\documentclass{WileyMSP-template}

\usepackage{xcolor}

\begin{document}

\pagestyle{fancy}
\rhead{\includegraphics[width=2.5cm]{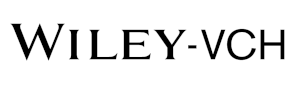}}

\title{Anisotropy-dependent decay of room temperature metastable skyrmions and a nascent double-$q$ spin texture in Co$_{8}$Zn$_{9}$Mn$_{3}$}

\maketitle


\author{Jonathan S.~White*}
\author{Victor~Ukleev}
\author{Le~Yu}
\author{Yoshinori~Tokura}
\author{Yasujiro~Taguchi}
\author{Kosuke Karube*}


\begin{affiliations}
Dr.~Jonathan S. White, Dr.~Victor Ukleev, Dr.~Le Yu\\
Laboratory for Neutron Scattering and Imaging (LNS), PSI Center for Neutron and Muon Sciences, Paul Scherrer Institut (PSI), CH-5232 Villigen PSI, Switzerland\\
Email Address: jonathan.white$@$psi.ch\\[1em]

Dr.~Victor Ukleev\\
Helmholtz-Zentrum Berlin für Materialien und Energie, Hahn-Meitner-Platz 1, D-14109, Berlin, Germany\\[1em]

Dr.~Le Yu\\
Laboratory for Ultrafast Microscopy and Electron Scattering (LUMES), Institute of Physics, \'Ecole Polytechnique F\'ed\'erale de Lausanne (EPFL), CH-1015 Lausanne, Switzerland\\[1em]

Dr.~Le Yu\\
Laboratory of Nanoscale Magnetic Materials and Magnonics (LMGN), Institute of Materials, \'Ecole Polytechnique F\'ed\'erale de Lausanne (EPFL), CH-1015 Lausanne, Switzerland\\[1em]

Prof.~Dr.~Yoshinori Tokura, Dr.~Yasujiro Taguchi, Dr.~Kosuke Karube\\
RIKEN Center for Emergent Matter Science (CEMS), Wako, 351-0198, Japan\\
Email Address: kosuke.karube$@$riken.jp\\[1em]

Prof.~Dr.~Yoshinori Tokura\\
Department of Applied Physics, The University of Tokyo, Tokyo, 113-8656, Japan\\[1em]

Prof.~Dr.~Yoshinori Tokura\\
Tokyo College, The University of Tokyo, Tokyo, 113-8656, Japan
\end{affiliations}



\begin{abstract}
Chiral cubic Co-Zn-Mn magnets exhibit diverse topological spin textures, including room-temperature skyrmion phases and robust far-from-equilibrium metastable states. Despite recent advances in understanding metastable skyrmions, the interplay between compositional disorder and varying magnetic anisotropy on the stability and decay of metastable textures, particularly near room temperature, remains incompletely understood. In this work, the equilibrium and metastable skyrmion formation in Co$_{8}$Zn$_{9}$Mn$_{3}$ is examined, revealing transformations between distinct metastable spin textures induced by temperature and magnetic field. At room temperature, the decay dynamics of metastable skyrmions exhibits a strong dependence on magnetic anisotropy, showcasing a route towards tailoring relaxation behavior. Furthermore, a nascent double-$q$ spin texture, characterized by two coexisting magnetic modulation vectors $q$, is identified as a minority phase alongside the conventional triple-$q$ hexagonal skyrmion lattice. This double-$q$ texture can be quenched as a metastable state, suggesting both its topological character, and its role as a potential intermediary of metastable skyrmion decay. These findings provide new insights into the tunability of equilibrium and metastable topological spin textures via chemical composition and magnetic anisotropy, offering strategies for designing materials with customizable and dynamic skyrmion properties for advanced technological applications.
\end{abstract}


\section{Introduction}
In materials science and technology, many applications are premised on the controlled decay of metastable states, where the transitions between states enable specific functionalities~\cite{Zha18,Pes23,Cui24}. For example, in pharmaceutical materials, the controlled decay of metastable compounds facilitates drug delivery and enhanced-performance therapeutic treatments~\cite{Sin04,Men15,Zha21,Wan23}. Energy storage concepts benefit from the release of energy in metastable materials~\cite{Raj24}, while in data storage, the controlled metastability of memory states allows for memory devices where information may be stored securely and then gradually erased, improving both data retention and security~\cite{Hui23,Yoo23,Naz24}. Magnetic skyrmions — topologically protected spin configurations with integer topological charges — readily exist as metastable states within magnetic crystals~\cite{Kan17,Tok21}, and are promising for diverse applications ranging from spintronics~\cite{Jia15,Mar21}, to magnonics~\cite{Gar17,Tak21}, to sensors~\cite{Zha15,Wie16}. Their controlled decay into trivial states, such as spirals, could be harnessed for example in sensors that depend on a dynamic in-situ modification of their sensitivity.\\[1em]

More fundamentally, studies of metastable skyrmions address questions concerning the energetics of topological phase transitions~\cite{Oik16,Yu18b,Bir19,Des20,Kar20}, out-of-equilibrium phases~\cite{Kar16,Nak17,Kar20}, topological protection far-from-equilibrium~\cite{Kag17b,Whi18,Ukl22,Tru23}, and individual skyrmion control~\cite{Sam13,Yua16,Fer17,Roz17}. Recent work further emphasises the role of entropic corrections on metastable skyrmion lifetimes~\cite{Wil17,Des18,Pot23}. Despite the recent advances, further developments in the general understanding of metastable skyrmion stability and control remain crucial for unlocking their full potential for applications.\\[1em]

Metastable skyrmions have been widely studied in chiral cubic magnets such as the $B$20s~\cite{Rit13,Oik16,Nak17,Ban17,Kag17b,Mun10,Bau16,Kag17,Wil17,Bau18,Yu18b,Tru23,Lit24,Kal24}, Cu$_{2}$OSeO$_{3}$~\cite{Oka16,Ban17,Mak17,Whi18,Bir19,Bir21}, and Co$_{x}$Zn$_{y}$Mn$_{z}$ ($x$+$y$+$z$=20) compounds~\cite{Kar16,Kar17,Kar20}. In these systems, Dzyaloshinskii-Moriya interactions stabilize helical ground states, with skyrmion lattices (SkLs) appearing near $T_{c}$ under small applied fields. Thermal fluctuations play a key role in high temperature ($T$) SkL formation~\cite{Muh09,Kru18}, while metastable skyrmions are typically created by field-cooling through the equilibrium SkL phase~\cite{Oik16,Kar16,Kar17,Bir19,Kar20}. Experiments show the relaxation dynamics of metastable skyrmions follow a modified Arrehnius law~\cite{Oik16,Kar16,Kar17,Wil17,Bir19,Kar20}, with lifetimes strongly influenced by structural and magnetic disorder~\cite{Bir19,Kar20}. In non-stoichiometric compounds substitutional disorder hinders decay mechanisms involving the motion of Bloch-point singularities~\cite{Mil13,Bir19,Kar20,Bir21}, resulting in exceptionally long metastable skyrmion lifetimes compared to stoichiometric systems.\\[1em]

In previous work on Co$_{10-x/2}$Zn$_{10-x/2}$Mn$_{x}$, the metastable skyrmion decay across all of the $x$=0,2,4 compounds could be described by just a single Arrehnius scaling~\cite{Kar20}. This finding could be considered surprising since varying $x$ is known to affect a non-trivial variation of random structural and magnetic disorder~\cite{Nak19,Boc19}, and in turn the equilibrium phase and metastable state diagrams~\cite{Kar16,Kar17,Kar18,Kar20,Pre21,Whi22,Kar22}. While disorder strongly influences skyrmion relaxation dynamics, the energy landscape necessarily also depends on the magnetic properties such as anisotropy, as well as entropic effects more broadly. To date however, the extent to which this interplay can be tuned to control metastable skyrmion relaxation, particularly in single-composition samples, remains little explored even at an empirical level. Harnessing this interplay at room temperature, however, could hold significant potential for advancing applications based on metastable skyrmions.\\[1em]

To explore these effects we studied Co$_{8}$Zn$_{9}$Mn$_{3}$ with $T_c\sim$313~K, thus chosen for its equilibrium SkL phase slightly above room temperature that allows for studying metastable skyrmion decay at room temperature (280-296~K)~\cite{Tok15}. Using magnetometry and small-angle neutron scattering (SANS), we created metastable skyrmion states and investigated their stability and decay as functions of temperature and field. Distinct to the earlier work on Co$_{10-x/2}$Zn$_{10-x/2}$Mn$_{x}$ compounds~\cite{Kar16}, Co$_{8}$Zn$_{9}$Mn$_{3}$ displays a different Arrehnius scaling that is moreover dependent on the direction of the magnetic field with respect to the crystal. This finding reveals magnetic anisotropy to be an important, tunable contributor to the energy landscape mediating the relaxation dynamics at room temperature. The importance of magnetic anisotropy at high temperature further manifests in the form of evidence for a nascent double-$q$ phase that coexists with the normal hexagonal SkL near $T_{c}$. This double-$q$ structure, potentially a meron-antimeron lattice, lacks full stability in the bulk but it can also be quenched as a metastable state, suggesting its topological character. Overall, these findings provide empirical insights that hint at pathways for controlling both the formation of specific topological spin textures, as well as their metastable properties in Co-Zn-Mn compounds.\\[1em]

\section{Results and Discussion}
\subsection{Magnetic phases, state diagrams and a nascent double-$q$ order}
Like the Co$_{10-x/2}$Zn$_{10-x/2}$Mn$_{x}$ sister compounds, Co$_{8}$Zn$_{9}$Mn$_{3}$ crystallises in the $\beta$-Mn chiral cubic spacegroup $P4_{1}32$ or $P4_{3}32$, which are related by mirror symmetry. As shown in \textbf{Figure~\ref{fig:Fig1}}a, the atoms occupy two Wyckoff sites; 8$c$ with a three-fold rotation symmetry that is majority filled by Co atoms, and 12$d$ with a two-fold rotation symmetry on which Zn and Mn mostly reside. Long period chiral magnetism is carried by Co on the 8$c$ site~\cite{Ukl19,Ukl21}, while the 12$d$ site forms a hyper-kagome network. The latter provides a fertile ground for geometrically-frustrated antiferromagnetic (AFM) interactions of any Mn atoms present~\cite{Ukl21}. For increasingly Mn-rich content, the development of frustrated AFM interactions gradually overwhelms Co chiral magnetism, eventually suppressing long-range helical and skyrmion order entirely~\cite{Kar20,Whi22}. For Co$_{8}$Zn$_{9}$Mn$_{3}$, the modest Mn content randomly occupies positions on the 12$d$ site, inducing weak pinning of skyrmions that enhances their stability against thermal agitation and assists in the creation of metastable skyrmions.\\[1em]

To clarify the magnetic properties of our single crystals we performed SANS [Figure~\ref{fig:Fig1}b and c] and magnetometry [Figure~\ref{fig:Fig1}d and e] measurements (see Methods for details of our experiments, and also Note S1, Figure S1, S2, in the Supporting Information for further characterisation data). At 310~K and in the zero-field helical phase, the SANS pattern in Figure~\ref{fig:Fig1}b shows magnetic scattering intensity to be concentrated mostly into a pair of spots described by a single propagation vector $\textbf{q}$$\parallel$$[100]$ (note that for each $q$-vector, scattering appears at both $\pm q$). Weaker spots due to a less populated helical domain are also observed with $q$ near to, but not precisely aligned with $[010]$. We attribute both the inequivalent helical domain populations, and the deviation of the minority domain's $q$-vector from alignment to the expected $[010]$ direction, as due to residual, inhomogeneous sample strains that are sufficient to overcome the weak bulk magnetic anisotropy close to $T_c$~\cite{Yu18}.\\[1em]

By applying a magnetic field of $\mu_{0}H$$\parallel$$[001]$=36~mT at 310~K, the SANS pattern transforms into that characteristic for the triple-$q$ hexagonal skyrmion lattice (SkL) [Figure~\ref{fig:Fig1}c]. Twelve Bragg spots are observed because the SANS pattern comprises two types of hexagonal SkL domains that are related by a 90$^{\circ}$ rotation, and for which one of the $q$-vectors is aligned with either [100] (red arrows in Figure~\ref{fig:Fig1}c) or [010] (white arrows). The observation of a multidomain SkL with $q$-vectors aligned with $\langle100\rangle$ type directions is common in the Mn-containing Co-Zn-Mn compounds with $\mu_{0}H$$\parallel$$[001]$~\cite{Kar16,Kar17,Kar18,Kar20,Whi22}. Since $q=2\pi/\lambda$, where $\lambda$ is the real-space periodicity of the magnetic order, from the SANS data we determined $\lambda$=103(1)~nm for both helical and SkL phases at 310~K.\\[1em]

According to the high-temperature portion of the magnetic phase diagram shown in Figure~\ref{fig:Fig1}d, the SANS data shown in Figure~\ref{fig:Fig1}c corresponds to that from an equilibrium SkL, which is stable between 300-313~K. The purple arrow in Figure~\ref{fig:Fig1}d denotes the typical thermodynamic route, i.e. the field-cooling (FC) process, adopted for creating metastable skyrmion states. In analogy with earlier work on the Co$_{10-x/2}$Zn$_{10-x/2}$Mn$_{x}$, x=0-6 compounds, a conventionally slow FC rate of $\sim-$1~K.min$^{-1}$ via the equilibrium skyrmion phase is sufficient to create metastable skyrmions in Co$_{8}$Zn$_{9}$Mn$_{3}$. The entire metastable state diagram determined by a.c. magnetometry is shown in Figure~\ref{fig:Fig1}e. Robust metastable skyrmions persist over an enormous phase space; down to very low temperatures, up to the saturation field in positive field, and a significant range of negative field. The transformation of the metastable SkL between hexagonal and what we term as an easy-axis distorted coordination is also identified, and elucidated further by SANS in the next sections. Overall, the general structure of the equilibrium phase diagram and metastable state diagrams for Co$_{8}$Zn$_{9}$Mn$_{3}$ is similar to those found earlier for Co$_{8}$Zn$_{8}$Mn$_{4}$ with $T_{c}$$\sim$300~K~\cite{Kar16,Kar20}.\\[1em]

Next we provide SANS evidence for a unique aspect of the equilibrium phase diagram in Co$_{8}$Zn$_{9}$Mn$_{3}$, namely a nascent double-$q$ structure that co-exists with the standard hexagonal SkL in the low-$T$ part of the equilibrium skyrmion phase pocket. As evidence for this hitherto unique co-existence of multi-$q$ structures, \textbf{Figure~\ref{fig:Fig2}}a shows the SANS pattern obtained at 305~K and 32~mT after an initial ZFC from 326~K. Twelve Bragg reflections can be identified that remain generally consistent with a two domain hexagonal SkL, but the Bragg spots with $\textbf{q}$$\parallel$$[100]$ and $[010]$ are more intense compared with the other eight spots, suggesting the existence of an additional magnetic order in the sample. As drawn in Figure~\ref{fig:Fig2}b, the overall structure and intensity of the SANS pattern in Figure~\ref{fig:Fig2}a can be described according to the superposition of SANS patterns due to the three structures, namely two approximately equally populated domains of hexagonal SkL and an additional double-$q$ structure. We point out that the additional intensity of the spots with $\textbf{q}$$\parallel$$[100]$ and $[010]$ cannot be due to single-$q$ helices, since the helical domains present in zero-field are essentially suppressed with increasing field before the transition into either conical or skyrmion order. In addition, later we present experimental evidence that the once-created double-$q$ structure can also be quenched as metastable state.\\[1em]

To support the observation of the double-$q$ structure quantitatively, in Figure~\ref{fig:Fig2}c we present an analysis of the SANS intensity data. In accord with the inset schematic, we distinguish between total intensities observed within the distinct red and blue box regions of the SANS detector. The schematic is also plotted in Figure~\ref{fig:Fig2}b, to highlight that the red-box intensities are only sensitive to scattering from eight spots due to the hexagonal SkLs, while the blue box intensities have contributions from four hexagonal SkL spots and the four spots of the double-$q$ structure. Using the overall red-box intensity, we estimate the contribution of the hexagonal SkL to the total blue-box intensities and then subtract it to leave only the residual intensities from phases that are not the hexagonal SkL.\\[1em]

The main panel of Figure~\ref{fig:Fig2}c shows the $\mu_{0}H$-dependence of the red-box intensities and the residual intensities from the blue boxes. The data shown were collected at both 310~K and 305~K after ZFC. At both temperatures, the red-symbol curves (due to red-box intensities) show an initial $\mu_{0}H$-induced suppression of the residual helical signal captured in the boxes, before the onset of the typical intensity peak due to the equilibrium hexagonal skyrmion phase. The blue-symbol curves show the $\mu_{0}H$-dependence of the residual blue-box intensities, which also shows the initial suppression of helical signals at low fields. The important difference between the two temperatures emerges at higher fields; while the data at 310~K remains close to zero as the field increases, i.e. consistent with no double-$q$ structure, a weak but clear intensity peak develops at 305~K denoting the onset of the double-$q$ structure as a minority phase coexisting with the hexagonal SkL.\\[1em]

From all of our $\mu_{0}H$- and $T$-dependent SANS data, Figs.~\ref{fig:Fig2}d and e show the high-$T$ portions of the phase diagrams, respectively highlighting the stability range of the equilibrium hexagonal SkL [Figure~\ref{fig:Fig2}d, with a SkL domain visualised schematically in Figure~\ref{fig:Fig2}f] and the double-$q$ structure [Figure~\ref{fig:Fig2}e]. The double-$q$ structure is seen to be only stable within the equilibrium hexagonal SkL phase, and the low-$T$ portion at that. The SANS intensity of the double-$q$ structure is a maximum at slightly lower temperatures compared with the hexagonal SkL indicating a fundamentally distinct origin. Since the maximum intensity of the double-$q$ structure is observed to be only 12(3)~\% the value of the coexisting two-domain hexagonal SkL order at 305~K and 36~mT, the double-$q$ order is clearly only nascent in the crystal.\\[1em]

In terms of the microscopic nature of the minority double-$q$ phase, candidate structures include a double-$q$ square-coordinated lattice of skyrmions, or a square lattice of merons and antimerons with half-integer topological charges~\cite{Yi09,Lin15,Pup20,Hay21,Leo24}. Clear-cut evidence for the latter in Co$_{8}$Zn$_{9}$Mn$_{3}$ was reported from earlier LTEM studies of thin plate samples, where the meron-antimeron lattice was observed directly to exist as a distinct phase on the low-$T$ side of the hexagonal SkL phase~\cite{Yu18}. There it was argued that the stability of a meron-antimeron phase benefits not only from the shape anisotropy of the sample in thin plate form, but also a developing easy-plane anisotropy on decrease of $T$. The latter may also be relevant in the present case, with micromagnetic simulations presented in Note S2 further supporting this suggestion. While our SANS data do not allow us to unambiguously establish the microscopic nature of the double-$q$ order in the bulk sample, in view of the earlier LTEM work, our micromagnetic simulations, and the location of the nascent double-$q$ phase on the lower-$T$ side of the hexagonal SkL phase pocket, a meron-antimeron lattice [like that shown schematically in Figure~\ref{fig:Fig2}g] becomes a plausible candidate structure.\\[1em]

In this case, the existence of inhomogeneous residual strains suggested from our analysis of the zero field helical order, may play a role in stabilising the meron-antimeron lattice through local enhancement of the easy-plane anisotropy in certain regions of the sample. In this context, future SANS studies on single crystals that focus on the fine control easy-plane anisotropy — such as by precise composition variation or placing samples under uniaxial strain — could provide a route towards decisive tuning of the mutual stabilities of the near-degenerate hexagonal SkL and double-$q$ orders. We finally note that real-space insights into the spin textures of the mm-scale bulk samples studied here may be accessible through implementation of a recently developed SANS tomographic technique~\cite{Hen23}. In addition, another promising direction for direct imaging on quasi-bulk ($\sim$1~$\mu$m thick) samples may involve harnessing coherent x-ray magnetic diffraction techniques, such as x-ray ptychography, as recently demonstrated for chiral magnets~\cite{Nee24}.

\subsection{SANS observations of metastable skyrmions}
\subsubsection{Temperature-dependence}
Next we turn to SANS data collected following the thermodynamic route outlined in \textbf{Figure~\ref{fig:Fig3}}a. The route starts with an initial FC process in $\mu_{0}H$$\parallel$$[001]$=36~mT from the equilibrium SkL phase at 310~K to 10~K, followed by removal of the field at 10~K and a zero-field warming (ZFW) back to 310~K. SANS patterns collected under selected conditions along the route are shown in Figure~\ref{fig:Fig3}b. We observe that during FC, the once-created equilibrium hexagonal SkL persists clearly as a robust metastable SkL state at 150~K. Upon further cooling, the twelve-spot pattern transforms abruptly to a square one with diffraction spots having $\textbf{q}$$\parallel$$[100]$ or $[010]$ at 10~K. After removal of the field and subsequent ZFW, we clearly observe the transformation from the low temperature fourfold pattern back to the twelve spot pattern characteristic of the metastable hexagonal SkL by 150~K. Upon further heating towards $T_c$, the metastable hexagonal SkL phase eventually decays into the equilibrium helical phase by $\sim$275~K. Importantly, the reemergence of the metastable hexagonal SkL by 150~K during ZFW confirms the preservation of topological charge at the transitions between the metastable hexagonal SkL states and the low temperature state. This strongly suggests that the magnetism underlying the fourfold SANS pattern at low $T$ remains rich in topological charge.\\[1em]

To analyse the SANS data more quantitatively, we followed the same approach as described in the previous section, defining distinct regions of the SANS detector [inset Figure~\ref{fig:Fig3}c] for which the propagation vectors [Figure~\ref{fig:Fig3}c] and SANS intensities [Figure~\ref{fig:Fig3}d] for the different types of magnetic order could be determined. Thus while in the red-box regions the total intensity is only from hexagonal SkLs (and residual helical order), subtraction of hexagonal SkL contribution to the blue box intensities leaves residual intensities due to either the low $T$ square SANS pattern, the double-$q$ square lattice, or helices, dependent on the location around the thermodynamic route.\\[1em]

The analysis shown in Figure~\ref{fig:Fig3}c shows that during the FC process, the length of the $q$-vector close to $T_c$ varies only weakly until $\sim$100~K. On further cooling it sharply increases until $\sim$35~K, which covers the transition between the metastable hexagonal SkL and a low $T$ state that yields square SANS patterns, remaining constant down to 10~K. The length of the $q$-vector at 10~K corresponds to structures with $\lambda$=67(1)~nm. Such a low temperature enhancement of the $q$-vector length along the $\langle100\rangle$ directions is seen commonly amongst the Mn containing Co$_{10-x/2}$Zn$_{10-x/2}$Mn$_{x}$, x=2-6 compounds, and understood to be driven by a cooperative interplay between developing AFM Mn-Mn spin correlations and enhanced magnetic anisotropy on cooling~\cite{Kar20,Ukl21,Whi22}.\\[1em]

Figure~\ref{fig:Fig3}d shows the total scattered SANS intensity from the different states throughout the thermodynamic route. Upon FC the intensity is initially dominated by the metastable hexagonal SkL, reaching a maximum near $\sim$180~K, before decreasing monotonically through the coordination transition and remaining finite at 10~K. A smaller contribution from the double-$q$ structure is also discernible during FC. Note that after the magnetic structure transition below $\sim$100~K, no clear Bragg spots due to a metastable hexagonal SkL are observed. Despite this, the finite red-box intensity suggests the persistence of orientationally-disordered metastable hexagonal SkL correlations, as will be argued in the next section. In contrast, the blue-box intensity sharply increases below 100~K due to the onset of the four strong spots characteristic of the low temperature fourfold SANS pattern. The ZFW sweep reveals a clear thermal hysteresis in the coordination transition observed on cooling. Notably on ZFW the intensity of the metastable hexagonal SkL does not fully recover to its value observed on FC, which we attribute due to a comparatively increased level of magnetic disorder.\\[1em]

The SANS observations of the FC process reported here for Co$_{8}$Zn$_{9}$Mn$_{3}$ are qualitatively similar to those observed earlier in Co$_{8}$Zn$_{8}$Mn$_{4}$~\cite{Kar16,Kar20}. As discussed for the latter compound, and also relevant here, the coordination transition between the metastable hexagonal SkL and the structure underlying the low temperature fourfold symmetric SANS pattern cannot be a straightforward hexagonal-to-square SkL transition, since the total topological charge (i.e. number of skyrmions) of the two structures cannot be preserved in the face of the enhancement of $q$ at low $T$s.\\[1em]

Instead, and as suggested by our micromagnetic simulations (see Note S2, Figure~S3, in the Supporting Information), the starting topological charge of the hexagonal SkL can be preserved through the transition if the metastable hexagonal SkL phase-separates at low temperatures into regions containing close-packed circular metastable skyrmions, and other regions where skyrmions distort along the direction of easy-axis magnetic anisotropy and form domains of elongated skyrmions. Such a scenario can explain the low temperature fourfold symmetric SANS pattern as constructed from scattering due to domains of skyrmions elongated along the easy axes such that $\textbf{q}$ is aligned mainly with $[100]$ or $[010]$. In addition, the persistent regions of orientationally-disordered yet close-packed circular metastable skyrmions can explain the finite SANS intensity observed away from $[100]$ and $[010]$ directions (i.e. the red-box intensities) at low temperature. Such a low $T$ deformation of metastable skyrmions is suggested as possible from earlier LTEM studies on Co$_{8}$Zn$_{8}$Mn$_{4}$~\cite{Mor17} and Co$_{8.5}$Zn$_{7.5}$Mn$_{4}$~\cite{Nag19} thin plate samples, and a related picture is likely relevant here in the present bulk Co$_{8}$Zn$_{9}$Mn$_{3}$ sample. Our micromagnetic simulation of the subsequent field removal and recovery of a closed-packed lattice of circular skyrmions in the ZFW process is also qualitatively consistent with the SANS observations (see Note S2, in the Supporting Information).

\subsubsection{Field-stability at 10~K}
Next we turn to the magnetic field stability of the FC states at 10~K following the different thermodynamic routes shown in \textbf{Figure~\ref{fig:Fig4}}a. Three different routes are explored: i) a field-increasing scan at 10~K after an initial ZFC from 326~K, ii) a field-increasing scan at 10~K after FC in $\mu_{0}H$$\parallel$$[001]$=36~mT from the equilibrium SkL phase at 310~K, and iii) a field-decreasing scan at 10~K after the same FC from 310~K.\\[1em]

Figure~\ref{fig:Fig4}b shows SANS patterns obtained at selected fields during the field-increasing sweep at 10~K after ZFC. Since this thermodynamic route avoids the equilibrium SkL phase entirely, the four strong clear spots with $\textbf{q}$$\parallel$$[100]$ or $[010]$ at 10~K constitute multiple domains of topologically-trivial helical order. As seen in Figure~\ref{fig:Fig4}b, increasing the magnetic field leads to a gradual suppression of scattering due to helical order as the system is driven into the (unobserved) conical phase. Complementary field-swept a.c. susceptibility data after ZFC are presented in Figure~\ref{fig:Fig4}e showing a symmetric field-dependence for both field-increasing and field-decreasing scans. The field-dependence of the SANS helical phase intensity after ZFC is shown in Figure~\ref{fig:Fig4}f. The intensity was determined as that located within the blue-box regions of the detector defined by the inset schematic, and it is observed to fall monotonically on the approach to the helical-conical transition.\\[1em]

Figure~\ref{fig:Fig4}c shows selected SANS patterns collected during the field-increasing scan after FC. Distinct to the ZFC process [Figure~\ref{fig:Fig4}b], once the four strong spots of the square SANS pattern are suppressed by the field, a residual ring of magnetic scattering intensity remains that is observed that survives until saturation. Since this ring of intensity is only fully revealed in field-sweeping after a FC protocol, it most likely corresponds to orientationally-disordered hexagonal correlations of metastable, circular skyrmions. Figure~\ref{fig:Fig4}d shows SANS patterns collected during the field-decreasing scan after FC, showing that reversing the field leads to a monotonic suppression of the four spot pattern for increasingly negative field, with no discernible intensity ring due to persistent metastable skyrmions observed.\\[1em]

In the corresponding a.c. susceptibility data [Figure~\ref{fig:Fig4}e], an inflection peak is observed in the field-increasing scan done after FC, whereby the metastable elongated skyrmion state is suppressed (the blue triangle near 200~mT), well before the suppression of the metastable hexagonal SkL correlations that survive until saturation (the blue triangle near $\sim$400~mT). When sweeping the field negative after FC, the blue triangle near $\sim$-190~mT denotes an anomaly indicating the transition out of the metastable elongated skyrmion state and into a helical phase as the field becomes increasingly negative. These features in a.c. susceptibility are backed up by the analysis of the SANS intensities shown in Figure~\ref{fig:Fig4}f. Following the same SANS analysis approach as described in the previous sections, we observe clearly that increasing the field after FC suppresses the intensity due to the four strong spots (the blue-box intensities) to zero by $\sim$200~mT, while the intensity due to metastable hexagonal SkL correlations remains robust to saturation (as determined from the red-box intensities). Upon sweeping the field negative after FC, kinks in the SANS intensities near $\sim$-180~mT in the field-dependence of both the red and blue box intensities denote the destabilisation of all metastable skyrmion states and their transition to trivial helices.\\[1em]

The field-sweeping behaviour described here is broadly consistent with our micromagnetic simulations (see Note S2, Figure~S4 in the Supporting Information), although the metastable skyrmion-to-helical transition suggested experimentally on sweeping the field negative is not reproduced. Despite this, the picture suggested from our data and simulations is that the distorted nature of metastable skyrmions at low temperatures makes them less robust to magnetic field in comparison to their co-existing counterparts that remain circular. This is presumably because the more symmetric distribution of magnetic moment - and hence topological charge - of circular skyrmions makes them comparatively more robust to external perturbation. In contrast, in-plane elongated skyrmions, which map to a description of short helical strings, will instead display local imperfections and a spatially diluted topological charge, leading to their increased susceptibility to destabilisation when exposed to an external field.

\subsubsection{Field-stability at 240~K}
Next we consider the magnetic field stability of the magnetic states at 240~K, which is closer to room temperature, and far from the thermally-driven low temperature transition between metastable skyrmion states. \textbf{Figure~\ref{fig:Fig5}}a shows the four different thermodynamics routes we explored in our SANS experiments: i) a field-increasing scan at 240~K after an initial ZFC from 326~K, ii) a field-increasing scan at 240~K after FC in $\mu_{0}H$$\parallel$$[001]$=36~mT from the equilibrium skyrmion phase at 310~K, iii) a field-decreasing scan at 240~K after the same FC from 310~K, and iv) a field-increasing scan done after reaching -32~mT in a preceding field-decreasing scan, all done after an initial FC from 310~K.\\[1em]

Starting with ZFC, selected SANS patterns from the subsequent field-increasing sweep at 240~K are shown in Figure~\ref{fig:Fig5}b. The SANS data are analogous to that seen at 310~K [Figure~\ref{fig:Fig1}b], evidencing the existence of unequally populated helical domains, the intensity of which are monotonically suppressed with increasing field as the sample transitions to the conical phase. The relevant field-swept a.c. susceptibility data are shown in Figure~\ref{fig:Fig5}f, revealing a symmetric field-dependence for both field-increasing and field-decreasing scans, and standard features associated with the helical-to-conical and conical-to-saturated phase transitions. Concomitant with the low-field dependence of the susceptibility data, Figure~\ref{fig:Fig5}g shows the gradual field-induced suppression of the helical phase SANS intensity, as evaluated on the SANS detector within the blue-box sectors shown in the inset schematic of Figure~\ref{fig:Fig5}g.\\[1em]

Figure~\ref{fig:Fig5}c shows selected SANS patterns collected during the field-increasing scan at 240~K after FC. Directly after FC we observe a clear twelve spot SANS pattern due to the superposed metastable states of multidomain hexagonal SkL and minority double-$q$ structure. Upon increasing the field, the metastable multi-$q$ structures persist for the majority of the field range, eventually becoming increasingly disordered to form an intensity ring just before the transition to the saturated phase. The corresponding field-dependent a.c. susceptibility [Figure~\ref{fig:Fig5}f] varies only smoothly until saturation with no clear anomalies denoting magnetic transitions, thus indicating the robustness of the metastable states. For the field-decreasing scan at 240~K after FC, the selected SANS patterns [Figure~\ref{fig:Fig5}d] shows the suppression of the metastable hexagonal SkL by small negative fields. In addition, as the intensity from the hexagonal SkL falls with increasingly negative field, the diffraction spots with $\textbf{q}$$\parallel$$[100]$ or $[010]$ become more visible, denoting the reemergence of helical-type domains. Eventually all intensity disappears for sufficiently strong negative fields, as the system transitions to the conical phase, this being consistent with that suggested from the negative field-sweep a.c. susceptibility data shown in Figure~\ref{fig:Fig5}f.\\[1em]

Figure~\ref{fig:Fig5}g shows the field-sweeping dependence of the SANS intensities for the different magnetic states created after FC to 240~K. Following the same analysis approach as described in previous sections, we see that the red-box intensities due to metastable hexagonal SkLs remains robust with positive field-sweeping before falling on the approach to saturation, while with negative field-sweeping, the associated intensity falls more quickly. From the concomitant analysis of the blue-box intensities, we observe the residual intensity due to the metastable double-$q$ lattice (blue squares in Figure~\ref{fig:Fig5}g) also survives until saturation, this providing further evidence as to its multi-$q$ nature with likely topological properties. For negative field-sweeping, the additional intensity initially increases, which we connect to the destabilisation of the double-$q$ structure into helices [c.f. Figure~\ref{fig:Fig5}d], before it falls sharply to zero at the helical-to-conical transition expected for sufficiently large negative fields.\\[1em]

Finally, we confirmed that the double-$q$ structure unwinds into helices under modest negative field sweeping by exploring the fourth thermodynamic route outlined in Figure~\ref{fig:Fig5}a. Namely, after performing the standard FC to 240~K and an initial negative field-sweep to -32~mT, the field was then reversed again and swept positive until saturation. Figure~\ref{fig:Fig5}e shows that in the final positive field-sweep, the intensity of the four more intense spots with $\textbf{q}$$\parallel$$[100]$ and $[010]$ at 0~mT are suppressed quickly with field, so that by 90~mT, only the metastable hexagonal SkL is clearly visible, with this state persisting until saturation. This picture is supported by the quantitative analysis of the SANS intensities in the final field-increasing sweep highlighted in Figure~\ref{fig:Fig5}h. After the partial field reversal until -32~mT, the red-box intensity due the remaining metastable hexagonal SkL is suppressed with increasing field, falling to zero on the approach to saturation. An anomaly in the red symbol data is seen in the positive field range where the residual helical intensity captured by the blue-boxes also fall quickly to zero. The sharp fall to zero of the helical intensity thus denotes the normal helical-to-conical phase transition, and confirms that the metastable double-$q$ lattice decays into helices upon the initial negative field-sweep to -32~mT after FC. The corresponding anomaly seen at the helical-to-conical transition in the red symbol data is thus less likely to represent an aspect of the metastable hexagonal SkL, but rather corresponds to the disappearance of a contribution to the red-boxes of disordered helices emerge along the explored thermodynamic route as they transition to a conical phase.

\subsection{Anisotropy-dependent decay of metastable skyrmions at room temperature}
The SANS and magnetometry data presented so far have revealed the existence of metastable skyrmion states in Co$_{8}$Zn$_{9}$Mn$_{3}$ under $\mu_{0}H$ and $T$ conditions where the metastable state lifetimes are practically infinite. Next we discuss time-dependent magnetometry measurements of metastable skyrmion decay, which at room temperature takes place on timescales ranging from months to years, these being accessible for experiments and relevant for putative applications.\\[1em]

\textbf{Figure~\ref{fig:Fig6}}a shows the high-$T$ portion of the $\mu_{0}H$ versus $T$ phase diagram for $\mu_{0}H$$\parallel$$[100]$. The time-dependence of the ac susceptibility $\chi'(t)$ was measured at fixed values of $\mu_{0}H$ and $T$ after following a FC procedure (purple arrow) to the measurement $T$s indicated by circle symbols (see Note~S4 and Figure~S6 in the Supporting Information for further detail on how the optimum quenching field of 36~mT for these measurements was determined). Time-dependent datasets at each $T$ were each preceded by a fresh FC. From these data, we present the quantity of normalised ac susceptibility $\chi'{_N}\equiv \left[\chi'({\infty})-\chi'(t)\right]/\left[\chi'({\infty})-\chi'(0)\right]$ as functions of both time and $T$ in Figure~\ref{fig:Fig6}b. Here $\chi'(0)$ and $\chi'(\infty)$ respectively correspond to the susceptibilities of the starting metastable skyrmion state, and the competing equilibrium conical phase into which the metastable skyrmion state decays, such that $\chi'{_N}=1$ at $t=0$, and $\chi'{_N}=0$ as $t\rightarrow\infty$. As seen in Figure~\ref{fig:Fig6}b, at each $T$ we observe a slow, yet clear decay of the metastable skyrmion state into the equilibrium conical phase, with relaxation times generally increasing as the $T$ is lowered, and estimated to vary from 10$^{7}$-10$^{8}$~s (several months to several years). Figs.~\ref{fig:Fig6}c and d show the analogous phase diagram and time-dependent $\chi'{_N}$ data obtained similarly for $\mu_{0}H$$\parallel$$[110]$. At each measurement $T$, $\chi'{_N}$ falls more slowly with time for $\mu_{0}H$$\parallel$$[110]$ compared with $\mu_{0}H$$\parallel$$[100]$, suggesting magnetic anisotropy as an effective factor influencing the metastable skyrmion lifetime.\\[1em]

To substantiate more quantitatively the role of magnetic anisotropy on metastable skyrmion decay, we fitted each time-dependent $\chi'{_N}$ dataset with a stretched exponential $\textrm{exp}\{-(t/\tau)^{\beta}\}$, where $\tau$ describes the characteristic relaxation time, while $\beta$ values lie in the range 0.3-0.5 indicating the existence of a broad distribution of relaxation times. In Figure~\ref{fig:Fig6}e we present the fitted values of $\tau$ versus reduced temperature $T/T_{c}$ for both directions of $\mu_{0}H$ investigated in Co$_{8}$Zn$_{9}$Mn$_{3}$. For each field direction, the values of $\tau$ are observed to increase exponentially as $T$ falls. Therefore we follow the approach of earlier work~\cite{Oik16,Wil17,Kar20} and fit the data to a modified Arrhenius law $\tau=\tau_{0}\textrm{exp}(a(T_{c}-T)/T)$. Here $a$ provides a measure of the effective energy barrier for skyrmion decay (normalised to $k_{B}T_{c}$), while $\tau_0$ represents the intrinsic time-scale, or `attempt time', of the decay process in the absence of thermal activation.\\[1em]

\begin{table}
\caption{Summary of fitted relaxation parameters for metastable skyrmion states in various materials.}
    \label{tab:decaytable}
    \centering
  \begin{tabular}[htbp]{cccc}
   \hline\hline
        Material & a & $\tau_0 (s)$ & Ref. \\
        \hline
        Co$_{8}$Zn$_{9}$Mn$_{3}$, $\mu{_0}H$$\parallel$$[100]$ & 92 & 5670 & This work\\
        Co$_{8}$Zn$_{9}$Mn$_{3}$, $\mu{_0}H$$\parallel$$[110]$ & 101 & 54010 & This work\\
        (Co$_{0.5}$Zn$_{0.5}$)$_{20-x}$Mn$_{x}$, $x=0,2,4$, $\mu{_0}H$$\parallel$$[110]$ & 215 & 63 & \cite{Kar20}\\
        Cu$_{2}$OSeO$_{3}$ & 96 & 3 & \cite{Bir19}\\
        Cu$_{2}$OSeO$_{3}$ (2.5~\% Zn-doped) & 94 & 150 & \cite{Bir19}\\
        MnSi & 65 & 0.00027 & \cite{Oik16}\\
        \hline
  \end{tabular}

\end{table}

\textbf{Table~\ref{tab:decaytable}} shows that for the two field directions, the fitted values of $a$ are comparable, yet $\sim$10\% larger for $\mu_{0}H$$\parallel$$[110]$ compared with $\mu_{0}H$$\parallel$$[100]$. The difference between $\tau_{0}$ values is approximately an order of magnitude however, namely much longer for $\mu_{0}H$$\parallel$$[110]$ ($\sim$5$\times$10$^{4}$~s) compared with $\mu_{0}H$$\parallel$$[100]$ ($\sim$5$\times$10$^{3}$~s). The significant field-direction anisotropy in $\tau_{0}$ most likely reflects the role of magnetic anisotropy on metastable skyrmion stability. Since the preferred alignment of the ground-state helical spiral is parallel to the cubic axes, the $\langle100\rangle$ directions correspond to the magnetic easy-axes. Thus, for $\mu_{0}H$$\parallel$$[100]$, the natural co-alignment of the propagation direction of the competing conical phase, which is parallel to the field, with that favoured by the magnetic anisotropy, facilitates the comparatively quicker decay of metastable skyrmions.\\[1em]

In contrast, for $\mu_{0}H$$\parallel$$[110]$, the direction of the applied field and thus the conical phase modulation is misaligned with the magnetic easy-axis, leading to an increased resistance to spin texture transformation during decay. This misalignment likely not only increases the effective energy barrier $a$ for transitioning to the conical phase, it also prolongs the metastable skyrmion state by ‘trapping’ the skyrmions for a longer, leading to higher values of $\tau_{0}$. These observations show that in the range of room temperature, the stability of the metastable skyrmion states in Co$_{8}$Zn$_{9}$Mn$_{3}$ can be tuned by varying the mutual alignment or misalignment between the directions of the magnetic field and magnetic easy-axis. This picture is supported further by the analysis of time-dependent susceptibility data collected under different conditions shown in Note S5, Figures S7 and S8. The analysis confirms the once-created metastable skyrmions are consistently more stable for $\mu_{0}H$$\parallel$$[110]$ compared with $\mu_{0}H$$\parallel$$[100]$ across all measured field strengths and temperatures investigated. At the fixed room $T$ of 296~K, we find that the anisotropy in both the characteristic decay time $\tau$ and energy barrier increase as the field strength is raised. In addition, $T$-dependent measurements performed in a fixed field smaller than the optimum of 36~mT suggest that the nature of the competing equilibrium phase may influence decay anisotropy, particularly for $\mu_{0}H$$\parallel$$[110]$, where deviations from Arrhenius behavior seem to emerge on passing through the helical-conical transition at lower temperatures.\\[1em]

Next we use Table~\ref{tab:decaytable} to compare between the $a$ and $\tau_0$ values obtained from the present analysis on Co$_{8}$Zn$_{9}$Mn$_{3}$, and those obtained earlier from (Co$_{0.5}$Zn$_{0.5}$)$_{20-x}$Mn$_{x}$, $x=0,2,4$ with $\mu{_0}H$$\parallel$$[110]$~\cite{Kar20}, and other skyrmion-hosting materials. The values of $a$ are material-specific, yet intriguingly the values for Co$_{8}$Zn$_{9}$Mn$_{3}$ are smaller compared to the compared to previously investigated Co-Zn-Mn compounds. A contributing reason for this may be the reduced skyrmion size in Co$_{8}$Zn$_{9}$Mn$_{3}$ of 103(1)~nm compared with at least 110~nm or greater in (Co$_{0.5}$Zn$_{0.5}$)$_{20-x}$Mn$_{x}$, $x=0,2,4$. Another influence may lie with the nascent double-$q$ state hitherto unique to Co$_{8}$Zn$_{9}$Mn$_{3}$. The proximity of this nearly degenerate state to the coexisting metastable SkL at room $T$ may provide additional relaxation pathways, and serve as an intermediate state that lowers the effective energy barrier $a$, thus influencing the $T$-dependent decay dynamics. Future spatio-temporal-resolved real-space imaging studies could provide insight into this suggestion that competing states putatively participate in the decay process.\\[1em]

Concerning values of $\tau_0$, Table~\ref{tab:decaytable} shows that larger values are clearly observed for materials displaying structural disorder related to partial chemical substitution and randomness in site occupancies. Considering (Co$_{0.5}$Zn$_{0.5}$)$_{20-x}$Mn$_{x}$, $x=0,2,4$, earlier work found both $a$ and $\tau_{0}$ to be essentially independent of $x$ [shown by the fit in Figure~\ref{fig:Fig6}e]~\cite{Kar20}, from which it was suggested that the decay was mainly governed by the inherent structural disorder that is always present. Co$_{8}$Zn$_{9}$Mn$_{3}$ could be expected to display a similar level of structural disorder to the $x=0,2,4$ compounds; however, the fitted values of $\tau_{0}$ for both field directions depart clearly from the scaling observed in those compositions. Our findings thus suggest that structural disorder alone does not account for metastable skyrmion decay in Co$_{8}$Zn$_{9}$Mn$_{3}$. This is particularly the case since magnetic indicators of disorder, such as broadening of the magnetic SANS peaks or evidence of subtle features in magnetic susceptibility, do not show a significant difference compared with the $x=0,2,4$ compounds. Instead, the observed field-direction anisotropy in metastable skyrmion relaxation shows magnetic anisotropy also governs the energy landscape near room temperature which, in concert with entropic effects more generally~\cite{Wil17,Des18,Bir21,Cri24}, conspire to generate a tunable distribution of accessible relaxation pathways. Building on this insight, further investigations are needed to develop a more precise understanding of the materials factors governing $\tau_{0}$ and $a$ in Co-Zn-Mn compounds, in order that these factors may be tailored for advanced applications.

\section{Conclusion}
Our study of Co$_8$Zn$_9$Mn$_3$ ($T_c$$\sim$313~K) has explored the rich interplay of chiral magnetism, disorder, anisotropy, and entropic effects that govern both the equilibrium phase diagram metastable skyrmion states. Using small-angle neutron scattering (SANS), we revealed the existence of a nascent double-$q$ structure - plausibly a meron-antimeron lattice - that coexists with the hexagonal skyrmion lattice (SkL) near $T_{c}$, and which we suggest to be stabilised by local enhancement of easy-plane anisotropy linked to residual strains. Such a sensitivity of the magnetic textures to variations in the microscopic interactions presents a promising avenue for the tuning of their mutual stability via precision control of composition, or through external parameters such as uniaxial strain.\\[1em]

Our observations of metastable skyrmion states created by field cooling (FC) reveal temperature- and magnetic field-dependent transitions, including distortions of the metastable hexagonal SkL driven by the concomitant development of both antiferromagnetic interactions and cubic anisotropy at cryogenic temperatures. At room temperature, we identified skyrmion relaxation dynamics to be anisotropic according to the orientation of the magnetic field. The observed decays follow a stretched exponential behaviour, with lifetimes spanning months to years which taken together highlight opportunities for precise control of the metastable skyrmion lifetimes on practical timescales.\\[1em]

More broadly, the interplay between chiral magnetism, disorder, magnetic anisotropy, and entropic effects provides a fertile platform for innovative skyrmion-based technologies. The exploitation of tunable skyrmion metastability at room temperature could enable diverse applications in computing, spintronics, and energy-efficient data storage. The robustness of metastable skyrmions against thermal agitation, coupled with their potential for transitions between energy-adjacent topological states, such as the double-$q$ structure uncovered here, makes them promising beyond advanced materials applications, potentially even for quantum information technologies~\cite{Psa21}, with the demonstrated room-temperature metastability of skyrmions in Co$_8$Zn$_9$Mn$_3$ emphasising the viability for ambient-condition applications. Future studies focussed on precise control of atomic structure, composition and strain engineering in Co-Zn-Mn compounds could provide pathways for systematically tuning all of skyrmion size, disorder, magnetic anisotropy, and relaxation dynamics, paving the way for material-specific properties that advance topological magnetism-based technologies.

\section{Experimental Section}
\threesubsection{Crystal growth and characterisation}\\
Following the approach reported elsewhere~\cite{Kar16,Kar17}, an ingot of Co$_8$Zn$_9$Mn$_3$ containing several single-crystalline grains was grown by the Bridgman method. Single crystalline grains were identified by x-ray Laue and cut out of the ingot. The extracted grains were cut along the (100), (010), and (001) planes to form rectangular-shaped crystals for SANS and magnetic susceptibility measurements.\\[1em]

\threesubsection{Small-angle neutron scattering (SANS) measurements}
The SANS experiments were performed using the SANS-I beamline located at the Swiss Spallation Neutron Source (SINQ), Paul Scherrer Institute (PSI), Switzerland. A 57.6~mg single crystal of Co$_8$Zn$_9$Mn$_3$ was mounted onto a purpose-built sample stick providing access to a wide temperature range from 2~K to beyond 350~K (and hence far above $T_c$$\sim$313~K). The sample stick was installed into a horizontal field cryomagnet at the beamline, and in a geometry where $\mu_{0}H$$\parallel$$[001]$$\parallel$$\textbf{k}_{i}$, where $\textbf{k}_{i}$ is the incident neutron wave vector. The chosen geometry is optimized for observing magnetic scattering intensity due to helical and skyrmion structures within the (001) plane perpendicular to the directions of both $\mu_{0}H$ and $\textbf{k}_{i}$.\\[1em]

For the SANS measurements, a velocity selector was used to select a neutron wavelength of 10~\AA with a 10~\% spread in full width at half-maximum. The incoming beam was collimated over a distance of 18~m before the sample, while the scattered neutrons were detected by a 1~m$^2$ two-dimensional multidetector placed 20~m behind the sample. The SANS data were collected by performing so-called rocking curve measurements, namely by collecting data at the cryomagnet and sample were rotated (or `rocked') around the vertical axis. Typically, data were collected over a range of rocking angle between the directions of $\mu_{0}H$ and $\textbf{k}_{i}$ of -15$^{\circ}$ to 15$^{\circ}$ scanned with a 2$^{\circ}$ step. At zero rocking angle, $\mu_{0}H$$\parallel$$\textbf{k}_{i}$. The typical time for a detector measurement at each angle was $\sim$10~s. To produce the SANS patterns presented in this work, the data obtained at all rocking angles are summed together. For the SANS data presented here, non-magnetic background scattering signals due to the crystal, sample environment and instrument were subtracted using further rocking curve data taken either above $T_c$, or at high field in the field-polarised phase. The SANS data reduction and analysis was performed using the GRASP software~\cite{Dew23}.\\[1em]

\threesubsection{Magnetometry measurements}
Direct-current (d.c.) magnetization data for single-crystal Co$_8$Zn$_9$Mn$_3$ were acquired using the vibrating-sample magnetometer mode of a superconducting quantum interference device magnetometer (Quantum Design MPMS3). Alternating-current (a.c.) susceptibility measurements were carried out using the a.c. measurement mode of the MPMS3. Both the static magnetic field and the a.c. excitation field (1~Oe) were applied along a cubic axis. The a.c. frequency was chosen to be 193~Hz.\\[1em]

\begin{figure}
\begin{centering}
    \includegraphics[width=\linewidth]{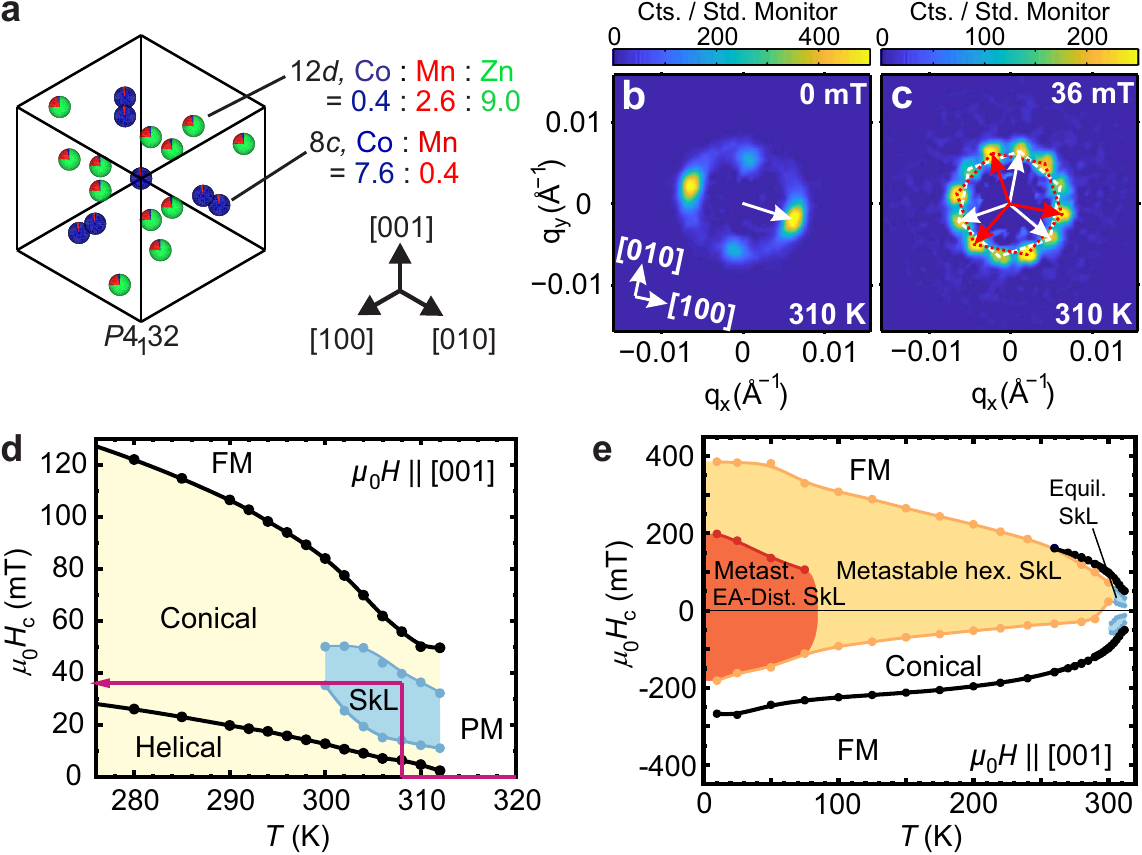}
    \caption{\textbf{Crystal structure, magnetic phases and state diagrams in Co$_{8}$Zn$_{9}$Mn$_{3}$}. \textbf{a} $\beta$-Mn-type cubic crystal structure of Co$_{8}$Zn$_{9}$Mn$_{3}$ viewed along the $[111]$ direction. The 12$d$ and 8$c$ Wyckoff sites are coloured according to their noted average occupation of Co (blue), Zn (green) and Mn (red). \textbf{b} and \textbf{c} show SANS data obtained at 310~K and \textbf{b} zero-field in the helical phase, and \textbf{c} 36~mT in the equilibrium skyrmion phase. The white and red arrows indicate unique propagation vectors $q$. For the skyrmion phase, two separate domains of triple-$q$ hexagonal SkLs are distinguished by red and white arrows. \textbf{d} High-temperature portion of the equilibrium magnetic phase diagram of Co$_{8}$Zn$_{9}$Mn$_{3}$ determined by a.c. susceptibility measurements for $\mu{_0}H$$\parallel$$[001]$. The equilibrium skyrmion phase (SkL) in blue exists between $T_c$$\sim$313~K and 300~K, while otherwise helical or conical states are stable. The purple arrow shows the measurement path of all the field-cooling (FC) measurements starting from 308~K, 36~mT in the equilibrium SkL phase. The resulting metastable state diagram determined by isothermal a.c. susceptibility measurements in field-increasing scans is shown in \textbf{e}. In \textbf{e}, the region of metastable SkL is enclosed by orange circle symbols. The region where a clear metastable hexagonal SkL is shaded light orange, while dark orange shading and red symbols denote the region where a metastable SkL state distorted by easy-axis anisotropy is stable.}
    \label{fig:Fig1}
    \end{centering}
\end{figure}

\begin{figure}
\begin{centering}
\includegraphics[width=0.5\linewidth]{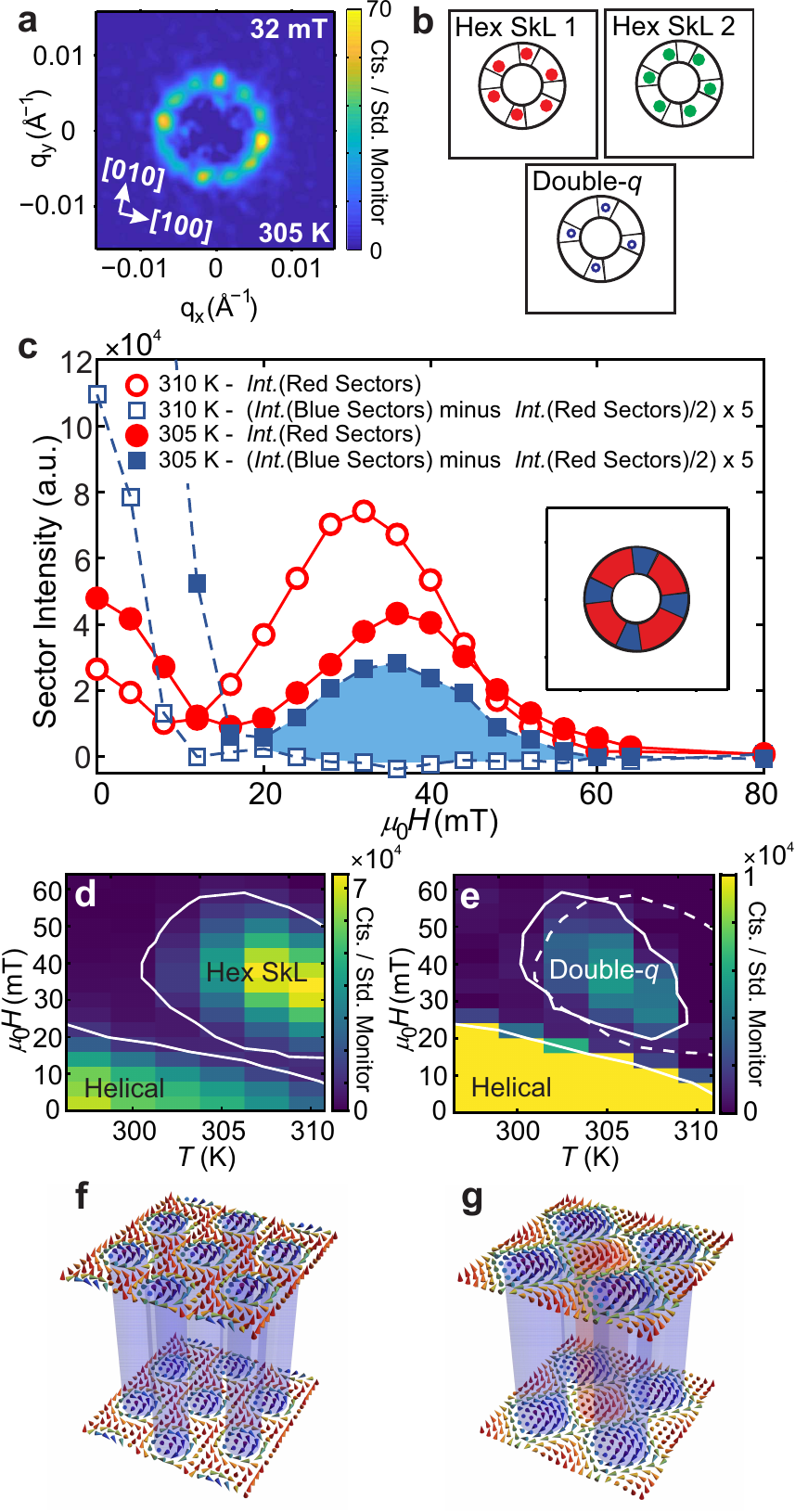}
    \caption{\textbf{Evidence for a double-$q$ state in Co$_{8}$Zn$_{9}$Mn$_{3}$}. \textbf{a} SANS data obtained at 305~K and 32~mT in the equilibrium skyrmion phase. \textbf{b} Schematic SANS patterns due hexagonal SkL domains (red and green spots), and the double-$q$ structure (empty blue circles). The superposition of the three patterns constitutes the overall diffraction pattern shown in \textbf{a}. In \textbf{b}, the overlaid annular sectors are relevant for analysis of SANS intensity data shown in \textbf{c}. \textbf{c} $\mu_{0}H$-increasing scans of the SANS intensities observed on the detector within the red and blue-coloured regions shown in the inset schematic. In the intermediate field region, the red (blue) symbol data show the extents of hexagonal SkL (double-$q$ structure) stability. The residual intensity detected in the blue boxes has been multiplied by five for improved visibility. The shaded blue region denotes the intensity observed from a double-$q$ structure (see the main text for detail of the analysis). \textbf{d} and \textbf{e} show the high-$T$ portion of the magnetic phase diagrams constructed from our SANS intensity analysis, highlighting the stability region of \textbf{d} the hexagonal SkL and \textbf{e} the double-$q$ structure. In \textbf{d} the solid white lines are constant intensity contours that denote the regions of stability of helical and hexagonal SkL phases. For clarity of phase stability, each contour is positioned at 1.2$\times$10$^{4}$ Cts./Std. Monitor, which for both helical and Hex SkL phases is 15~\% of their maximum intensity in the colormap. Similarly, in \textbf{e} the solid white contours denote the stability regions of the helical (1.2$\times$10$^{4}$ Cts./Std. Monitor) and double-$q$ structure (900 Cts./Std. Monitor), positioned again at 15~\% of the respective maximum intensities. The dashed white line corresponds to the stability region of the hexagonal SkL shown in \textbf{d}, indicating the double-$q$ structure is only observed to be stable in the presence of a coexisting hexagonal SkL. \textbf{f} and \textbf{g} respectively show schematics of a hexagonal SkL and a meron-antimeron lattice, the latter being a plausible candidate magnetisation texture for the double-$q$ structure.}
    \label{fig:Fig2}
    \end{centering}
\end{figure}

\begin{figure}
\begin{centering}
\includegraphics[width=\linewidth]{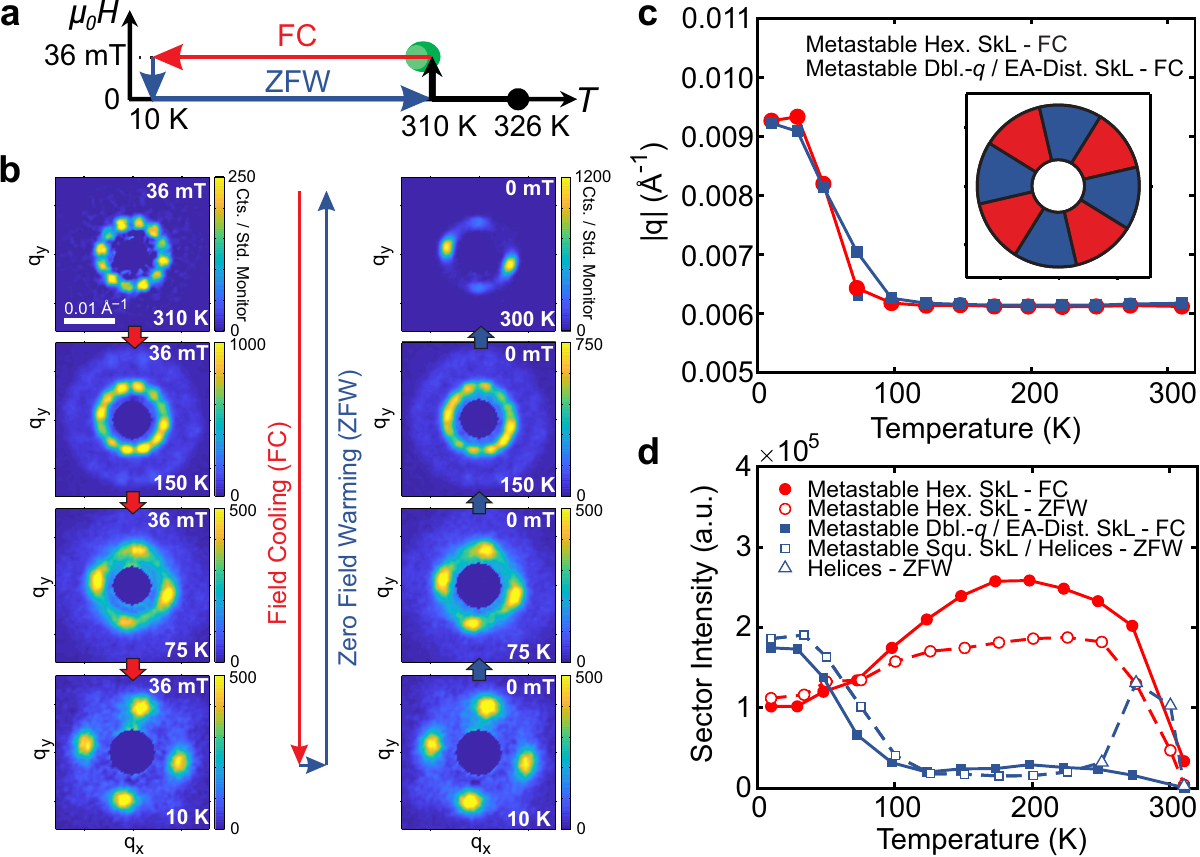}
    \caption{\textbf{SANS observations of metastable skyrmions in Co$_{8}$Zn$_{9}$Mn$_{3}$}. \textbf{a} Schematic illustration of the thermodynamic route over which SANS measurements were performed. The field-cooling (FC) and subsequent zero-field warming (ZFW) procedures are denoted by red and blue arrows, respectively. \textbf{b} SANS data collected at selected temperatures and fields during the FC-ZFW procedure outlined in panel \textbf{a}. The reciprocal space scale bar in the top-left panel applies to all SANS images. \textbf{c} Temperature-dependence of the propagation vectors $q$ of the metastable hexagonal skyrmion lattice (SkL) component (red circles) and the intensity allocated to be due to metastable double-$q$ or low $T$ skyrmion states giving a square SANS pattern (blue squares). The inset shows the regions of the SANS detector over which the propagation vectors and intensities (shown in panel \textbf{d)}) of the metastable hexagonal SkLs (red sectors), and the metastable double-$q$, metastable low $T$ skyrmions, or helices (i.e. the non-hexagonal SkL intensity captured in the blue sectors). Further details of the analysis are given in the main text. \textbf{d} Temperature-dependence of the SANS intensities from the metastable hexagonal SkL (circles), metastable double-$q$/low-$T$ skyrmions giving a square SANS pattern (square symbols), and helical order (triangles). Filled (empty) symbols denote the FC and ZFW procedures, respectively. In \textbf{c} and \textbf{d}, lines are guides for the eye.}
    \label{fig:Fig3}
    \end{centering}
\end{figure}

\begin{figure}
\begin{centering}
  \includegraphics[width=0.9\linewidth]{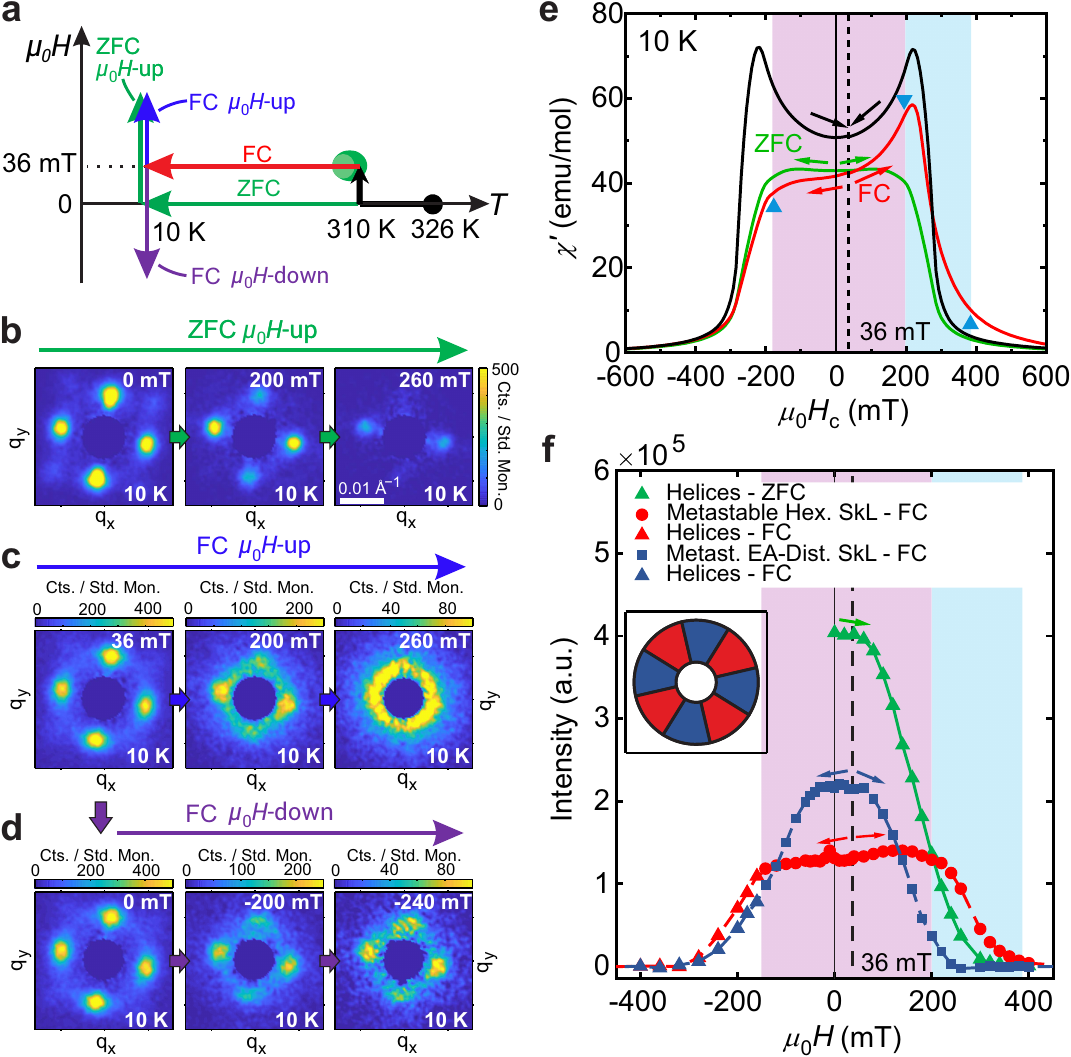}
    \caption{\textbf{Field stability of metastable skyrmions at 10~K}. \textbf{a} Schematic illustration of the routes in $\mu_{0}H$ and $T$ over which SANS measurements were performed. Green arrows denote the zero-field cool (ZFC) procedure from 326~K to 10~K, followed by SANS measurements at 10~K in a field-increasing scan (ZFC $\mu{_0}H$-up). Metastable SkL states were prepared at 10~K following field-cooling (FC) in 36~mT shown by the red arrow. Subsequently, SANS measurements were carried out for either field-increasing (blue arrow, FC $\mu{_0}H$-up) or field-decreasing (purple arrow, FC $\mu{_0}H$-down) scans. In \textbf{b}-\textbf{d}, SANS data are shown that were collected at selected fields during \textbf{b} the ZFC $\mu{_0}H$-up scan, \textbf{c} the FC $\mu{_0}H$-up scan, and \textbf{d} the FC $\mu{_0}H$-down scan. In the right-most panel of \textbf{b}, the reciprocal space scale bar is shown and applies to all SANS images in \textbf{b}-\textbf{d}. \textbf{e} Field-swept a.c. susceptibility data obtained at 10~K after a FC procedure. Field-swept data from 36~mT to +500mT or -500mT is shown by red lines, the return sweeps from +500mT or -500mT back to 36~mT are shown by black lines. Green lines show field-sweeping data obtained after ZFC. Blue triangle symbols denote either the boundaries or transitions of metastable skyrmion states. \textbf{f} Field-swept SANS intensity data obtained at 10~K after either an initial ZFC (green symbols), or instead a FC procedure (red and blue symbols). Intensity is allocated to be due to either helices after ZFC (green triangles), a metastable hexagonal SkL component (red circles), a residual helical component (red triangles), a metastable SkL that gives the square SANS pattern (blue squares) and a further residual helical component (blue triangles).  The inset schematic shows regions of the SANS detector that were used to determine the intensities allocated to the metastable states. (see the main text for further details).}
    \label{fig:Fig4}
    \end{centering}
\end{figure}

\begin{figure}
\begin{centering}
  \includegraphics[width=0.65\linewidth]{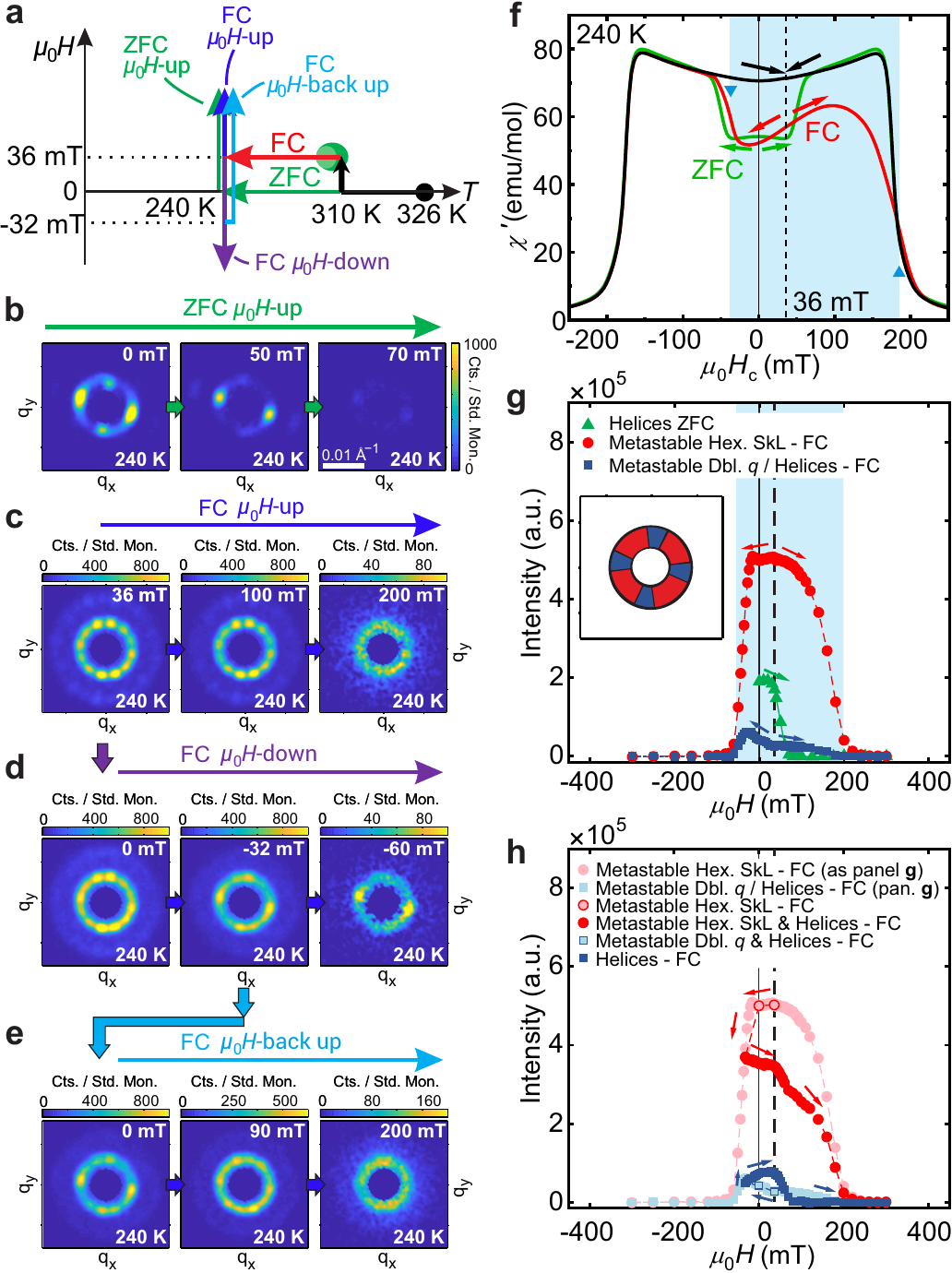}
    \caption{\textbf{Metastable skyrmions at 240~K}. \textbf{a} Schematic illustration of the routes in $\mu_{0}H$ and $T$ over which SANS measurements were performed. Green arrows denote the ZFC procedure from 326~K to 240~K, followed by SANS measurements at 240~K in a field-increasing scan (ZFC $\mu{_0}H$-up). Metastable SkL states were prepared at 240~K following the FC procedure in 36~mT shown by the red arrow. Subsequently, SANS measurements were carried out for either field-increasing (blue arrow, FC $\mu{_0}H$-up) or field-decreasing (purple arrow, FC $\mu{_0}H$-down) scans. A light blue arrow denotes a field-increasing scan done after reaching -32~mT in a preceding field-decreasing scan. Panels \textbf{b}-\textbf{e} show SANS data collected at selected fields during \textbf{b} the ZFC $\mu{_0}H$-up scan, \textbf{c} the FC $\mu{_0}H$-up scan, \textbf{d} the FC $\mu{_0}H$-down scan, and \textbf{e} the $\mu{_0}H$-back up scan initiated after the preceding field-decreasing scan. In the right-most panel of \textbf{b}, the reciprocal space scale bar is shown and applies to all SANS images in \textbf{b}-\textbf{e}. \textbf{f} Field-swept a.c. susceptibility data obtained at 10~K after FC. Field-swept data from 36~mT to +500mT or -500mT are shown by red lines, the return sweeps from +500mT or -500mT back to 36~mT are shown by black lines. Green lines show field-sweeping data obtained after ZFC. Blue triangle symbols denote either the boundaries or transitions of metastable skyrmion states. \textbf{g} Field-swept SANS intensity data obtained at 240~K after either the initial ZFC (green symbols), or a field-cooled (FC) procedure (red and blue symbols). Intensity is allocated to be due to either helices after ZFC (green triangles), a metastable hexagonal SkL component (red circles), or a component due the metastable double-$q$ order or helices (blue squares). The inset schematic shows regions of the SANS detector that were used to determine the intensities allocated to the different states. \textbf{h} Field-swept SANS intensity data obtained at 240~K upon the $\mu{_0}H$-sweep back up following an initial FC, field-decreasing sweep down to -32~mT. The data are described in detail in the main text.}
    \label{fig:Fig5}
    \end{centering}
\end{figure}

\begin{figure}
\begin{centering}
  \includegraphics[width=\linewidth]{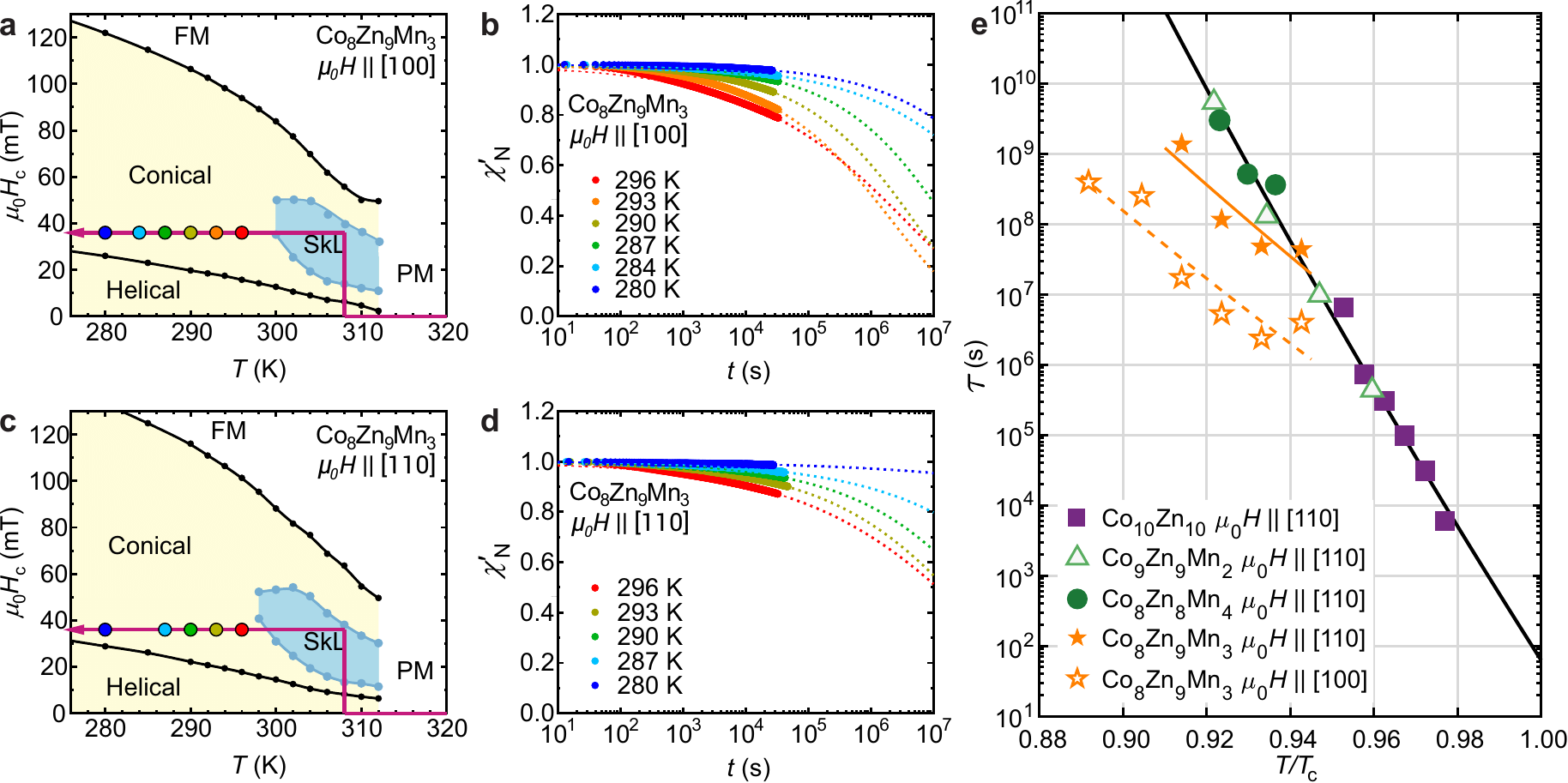}
    \caption{\textbf{Time- and temperature-dependent a.c. susceptibility measurements of metastable skyrmions in Co$_{8}$Zn$_{9}$Mn$_{3}$}. \textbf{a} The high temperature portion of the magnetic phase diagram of Co$_{8}$Zn$_{9}$Mn$_{3}$ determined by ac susceptibility ($\chi'$) for $\mu{_0}H$$\parallel$$[100]$. The purple arrow passing through the equilibrium skyrmion phase (shaded blue) denotes the route of field-cooling a metastable skyrmion state. Time-dependent measurements of $\chi'$ were done at the temperatures denoted by coloured circles. \textbf{b} Time-dependent measurements of the normalized ac susceptibility $\chi'{_N}$ (defined in the main text) at temperatures corresponding to those in panel \textbf{a}. The dotted lines denote the fit of the data at each temperature to a stretched exponential $\textrm{exp}\{-(t/\tau)^{\beta}\}$. \textbf{c} and \textbf{d} show results analogous to those for panels \textbf{a} and \textbf{b} but for $\mu{_0}H$$\parallel$$[110]$. \textbf{e} The fitted relaxation times $\tau$ of the metastable skyrmion state versus normalised temperature $T/T_{\textrm{c}}$ for Co$_{8}$Zn$_{9}$Mn$_{3}$. Solid and dashed orange lines are fits of the data using a modified Arrhenius law (see text for details). For comparison, we include additional data for (Co$_{0.5}$Zn$_{0.5}$)$_{20-x}$Mn$_{x}$, $x=0,2,4$ with $\mu{_0}H$$\parallel$$[110]$ taken from Ref.~\cite{Kar20}. A single fit of the modified Arrhenius law for these three compositions is denoted by the black line.}
    \label{fig:Fig6}
    \end{centering}
\end{figure}

\threesubsection{Demagnetization calibration}
Due to the shape difference between samples used in the magnetometry and SANS measurements their relative demagnetization factors are different. To present all data in this paper on a unified magnetic field scale, the magnetic field values for the magnetometry data are calibrated by $\mu_{0}H_c$ = $\alpha$$\times$$\mu_{0}H$. We determined coefficient $\alpha$=2.25 for data presented in Figures~\ref{fig:Fig1}d, \ref{fig:Fig6}a, and \ref{fig:Fig6}c, and $\alpha$=3 for Figures~\ref{fig:Fig1}e, \ref{fig:Fig4}e, \ref{fig:Fig5}f, S1b, S1c, S2, S6 all panels, S7 all panels, S8a, and S8c. For all the figures related to magnetometry measurements the calibrated field values are given with notation of $\mu_{0}H_c$.\\[1em]

\threesubsection{Micromagnetic simulation}
The SANS results for the temperature- and field-driven transformation of the metastable skyrmion lattice are compared with the results of Landau-Lifshitz-Gilbert (LLG) simulations performed using Mumax$^3$ package \cite{Van14}. Since precise estimates of micromagnetic parameters for Co$_{8}$Zn$_{9}$Mn$_{3}$ are not available in the literature and cannot be directly determined from our bulk data, we employed material parameters reported for Co$_8$Zn$_8$Mn$_4$ \cite{Ukl19,Hen22}. This choice is justified by the close similarities in the phase diagram and magnetic texture periodicities between Co$_8$Zn$_8$Mn$_4$ and Co$_{8}$Zn$_{9}$Mn$_{3}$, making the parameters of the former a reasonable approximation for understanding the behavior of the latter. The parameters used in our simulations are thus: saturation magnetization $M_s=350$\,kA/m, DMI constant $D=0.00053$\,J/m, magnetocrystalline cubic anisotropy $K_c = 5000$\,J/m$^3$, and exchange stiffness $A_{ex} = 9.2$\,pJ/m. The simulation geometry consisted of $2000\times2000\times1$ cells with the size of $5\times5\times200$\,nm$^3$.\\[1em]

\medskip
\textbf{Supporting Information} \par 
Supporting Information is available from the Wiley Online Library or from the author.

\medskip
\textbf{Acknowledgements} \par 
Financial support is gratefully acknowledged from the Swiss National Science Foundation (SNSF) via the Sinergia network `NanoSkyrmionics' (grant CRSII5\textunderscore171003), SNSF project grant No. 200021\textunderscore188707, JST CREST (Grant No. JPMJCR20T1 and JPMJCR1874), JST FOREST (Grant No. JPMJFR235R), JSPS Grant-in-Aids for Scientific Research (Grant no. 23K26534), and the RIKEN TRIP initiative (Many-body Electron Systems, Advanced General Intelligence for Science Program, and Fundamental Quantum Science Program). This work is based partly on experiments performed at the Swiss Spallation Neutron source SINQ, Paul Scherrer Institute, Villigen, Switzerland.

\medskip

%
\bibliographystyle{MSP}
\bibliography{893_bib}

\begin{thebibliography}{10}
\providecommand{\url}[1]{\texttt{#1}}
\providecommand{\urlprefix}{URL }

\bibitem{Zha18}
W.~Zhao, J.~Pan, Y.~Fang, X.~Che, D.~Wang, K.~Bu, F.~Huang,
\newblock \emph{Chemistry – A European Journal} \textbf{2018}, \emph{24}
  15942.

\bibitem{Pes23}
P.~Pesode, S.~Barve,
\newblock \emph{J. Eng. Appl. Sci.} \textbf{2023}, \emph{70} 25.

\bibitem{Cui24}
X.~Cui, Y.~Liu, Y.~Chen,
\newblock \emph{National Science Review} \textbf{2024}, \emph{11} nwae033.

\bibitem{Sin04}
D.~Singhal, W.~Curatolo,
\newblock \emph{Advanced Drug Delivery Reviews} \textbf{2004}, \emph{56} 335.

\bibitem{Men15}
F.~Meng, A.~Trivino, D.~Prasad, H.~Chauhan,
\newblock \emph{European Journal of Pharmaceutical Sciences} \textbf{2015},
  \emph{71} 12.

\bibitem{Zha21}
J.~Zhang, Q.~Shi, T.~Qu, D.~Zhou, T.~Cai,
\newblock \emph{International Journal of Pharmaceutics} \textbf{2021},
  \emph{610} 121235.

\bibitem{Wan23}
Y.~Wang, F.~Li, J.~Xin, J.~Xu, G.~Yu, Q.~Shi,
\newblock \emph{Molecules} \textbf{2023}, \emph{28}, 8.

\bibitem{Raj24}
C.~Raju, Z.~Sun, R.~Koibuchi, J.~Y. Choi, S.~Chakraborty, J.~Park, H.~Houjou,
  K.~Schmidt-Rohr, G.~G.~D. Han,
\newblock \emph{J. Mater. Chem. A} \textbf{2024}, \emph{12} 26678.

\bibitem{Hui23}
H.~Xia, S.~Xu, J.~Pei, R.~Zhang, Z.~Yu, W.~Zou, L.~Wang, C.~Liu,
\newblock \emph{IEEE Journal on Selected Areas in Communications}
  \textbf{2023}, \emph{41} 3573.

\bibitem{Yoo23}
C.~Yoon, G.~Oh, S.~Kim, J.~Jeon, J.~H. Lee, Y.~H. Kim, B.~H. Park,
\newblock \emph{NPG Asia Materials} \textbf{2023}, \emph{15} 33.

\bibitem{Naz24}
N.~P. Nazirkar, S.~Srinivasan, R.~Harder, E.~Fohtung,
\newblock \emph{AIP Advances} \textbf{2024}, \emph{14} 015155.

\bibitem{Kan17}
N.~Kanazawa, S.~Seki, Y.~Tokura,
\newblock \emph{Adv. Mater.} \textbf{2017}, \emph{29} 1603227.

\bibitem{Tok21}
Y.~Tokura, N.~Kanazawa,
\newblock \emph{Chem. Rev.} \textbf{2021}, \emph{121} 2857.

\bibitem{Jia15}
W.~J. Jiang, P.~Upadhyaya, W.~Zhang, G.~Q. Yu, M.~B. Jungfleisch, F.~Y. Fradin,
  J.~E. Pearson, Y.~Tsernovnyak, K.~L. Wang, O.~Heinonen, S.~G.~E. Te~Velthuis,
  A.~Hoffmann,
\newblock \emph{Science} \textbf{2015}, \emph{349} 283.

\bibitem{Mar21}
C.~H. Marrows, K.~Zeissler,
\newblock \emph{Appl. Phys. Lett.} \textbf{2021}, \emph{119} 250502.

\bibitem{Gar17}
M.~Garst, J.~Waizner, D.~Grundler,
\newblock \emph{Journal of Physics D: Applied Physics} \textbf{2017},
  \emph{50}, 29 293002.

\bibitem{Tak21}
R.~Takagi, M.~Garst, J.~Sahliger, C.~H. Back, Y.~Tokura, S.~Seki,
\newblock \emph{Phys. Rev. B} \textbf{2021}, \emph{104} 144410.

\bibitem{Zha15}
X.~Zhang, M.~Ezawa, Y.~Zhou,
\newblock \emph{Sci. Rep.} \textbf{2015}, \emph{5} 9400.

\bibitem{Wie16}
R.~Wiesendanger,
\newblock \emph{Nat. Rev. Mater.} \textbf{2016}, \emph{1} 16044.

\bibitem{Oik16}
H.~Oike, A.~Kikkawa, N.~Kanazawa, Y.~Taguchi, M.~Kawasaki, Y.~Tokura,
  F.~Kagawa,
\newblock \emph{Nat. Phys.} \textbf{2016}, \emph{12} 62.

\bibitem{Yu18b}
X.~Z. Yu, D.~Morikawa, T.~Yokouchi, K.~Shibata, N.~Kanazawa, F.~Kagawa,
  T.~Arima, Y.~Tokura,
\newblock \emph{Nat. Phys.} \textbf{2018}, \emph{14} 832.

\bibitem{Bir19}
M.~T. Birch, R.~Takagi, S.~Seki, M.~N. Wilson, F.~Kagawa,
  A.~\v{S}tefan\v{c}i\v{c}, G.~Balakrishnan, R.~Fan, P.~Steadman, C.~J. Ottley,
  M.~Crisanti, R.~Cubitt, T.~Lancaster, Y.~Tokura, P.~D. Hatton,
\newblock \emph{Phys. Rev. B} \textbf{2019}, \emph{100} 014425.

\bibitem{Des20}
L.~Desplat, C.~Vogler, J.-V. Kim, R.~L. Stamps, D.~Suess,
\newblock \emph{Phys. Rev. B} \textbf{2020}, \emph{101} 060403.

\bibitem{Kar20}
K.~Karube, J.~S. White, V.~Ukleev, C.~D. Dewhurst, R.~Cubitt, A.~Kikkawa,
  Y.~Tokunaga, H.~M. R\o{}nnow, Y.~Tokura, Y.~Taguchi,
\newblock \emph{Phys. Rev. B} \textbf{2020}, \emph{102} 064408.

\bibitem{Kar16}
K.~Karube, J.~S. White, N.~Reynolds, J.~L. Gavilano, H.~Oike, A.~Kikkawa,
  F.~Kagawa, Y.~Tokunaga, H.~M. R\o~nnow, Y.~Tokura, Y.~Taguchi,
\newblock \emph{Nat. Mater.} \textbf{2016}, \emph{15} 1237.

\bibitem{Nak17}
T.~Nakajima, H.~Oike, A.~Kikkawa, E.~P. Gilbert, N.~Booth, K.~Kakurai,
  Y.~Taguchi, Y.~Tokura, F.~Kagawa, T.~Arima,
\newblock \emph{Sci. Adv.} \textbf{2017}, \emph{3} 1602562.

\bibitem{Kag17b}
F.~Kagawa, H.~Oike, W.~Koshibae, A.~Kikkawa, Y.~Okamura, Y.~Taguchi,
  N.~Nagaosa, Y.~Tokura,
\newblock \emph{Nat. Commun.} \textbf{2017}, \emph{8} 1332.

\bibitem{Whi18}
J.~S. White, I.~\ifmmode \check{Z}\else \v{Z}\fi{}ivkovi\ifmmode~\acute{c}\else
  \'{c}\fi{}, A.~J. Kruchkov, M.~Bartkowiak, A.~Magrez, H.~M. R\o{}nnow,
\newblock \emph{Phys. Rev. Applied} \textbf{2018}, \emph{10} 014021.

\bibitem{Ukl22}
V.~Ukleev, D.~Morikawa, K.~Karube, A.~Kikkawa, K.~Shibata, Y.~Taguchi,
  Y.~Tokura, T.-h. Arima, J.~S. White,
\newblock \emph{Advanced Quantum Technologies} \textbf{2022}, \emph{5}, 11
  2200066.

\bibitem{Tru23}
B.~Truc, A.~A. Sapozhnik, P.~Tengdin, E.~V. Bostr\"{o}m, T.~Sch\"{o}nenberger,
  S.~Gargiulo, I.~Madan, T.~LaGrange, A.~Magrez, C.~Verdozzi, A.~Rubio, H.~M.
  R\o~nnow, F.~Carbone,
\newblock \emph{Adv. Mater.} \textbf{2023}, \emph{35} 2304197.

\bibitem{Sam13}
J.~Sampaio, V.~Cros, S.~Rohart, A.~Thiaville, A.~Fert,
\newblock \emph{Nature Nanotech.} \textbf{2013}, \emph{8} 839.

\bibitem{Yua16}
H.~Y. Yuan, X.~R. Wang,
\newblock \emph{Sci. Rep.} \textbf{2016}, \emph{6} 22638.

\bibitem{Fer17}
A.~Fert, N.~Reyren, V.~Cros,
\newblock \emph{Nature Reviews Materials} \textbf{2017}, \emph{2} 17031.

\bibitem{Roz17}
L.~R\'ozsa, K.~Palot\'as, A.~De\'ak, E.~Simon, R.~Yanes, L.~Udvardi,
  L.~Szunyogh, U.~Nowak,
\newblock \emph{Phys. Rev. B} \textbf{2017}, \emph{95} 094423.

\bibitem{Wil17}
J.~Wild, T.~N.~G. Meier, S.~P{\"o}llath, M.~Kronseder, A.~Bauer, A.~Chacon,
  M.~Halder, M.~Schowalter, A.~Rosenauer, J.~Zweck, J.~M{\"u}ller, A.~Rosch,
  C.~Pfleiderer, C.~H. Back,
\newblock \emph{Sci. Adv.} \textbf{2017}, \emph{3} 1701704.

\bibitem{Des18}
L.~Desplat, D.~Suess, J.-V. Kim, R.~L. Stamps,
\newblock \emph{Phys. Rev. B} \textbf{2018}, \emph{98} 134407.

\bibitem{Pot23}
M.~N. Potkina, I.~S. Lobanov, H.~J\'onsson, V.~M. Uzdin,
\newblock \emph{Phys. Rev. B} \textbf{2023}, \emph{107} 184414.

\bibitem{Rit13}
R.~Ritz, M.~Halder, C.~Franz, A.~Bauer, M.~Wagner, R.~Bamler, A.~Rosch,
  C.~Pfleiderer,
\newblock \emph{Phys. Rev. B} \textbf{2013}, \emph{87} 134424.

\bibitem{Ban17}
L.~J. Bannenberg, F.~Qian, R.~M. Dalgliesh, N.~Martin, G.~Chaboussant,
  M.~Schmidt, D.~L. Schlagel, T.~A. Lograsso, H.~Wilhelm, C.~Pappas,
\newblock \emph{Phys. Rev. B} \textbf{2017}, \emph{96} 184416.

\bibitem{Mun10}
W.~M\"unzer, A.~Neubauer, T.~Adams, S.~M\"uhlbauer, C.~Franz, F.~Jonietz,
  R.~Georgii, P.~B\"oni, B.~Pedersen, M.~Schmidt, A.~Rosch, C.~Pfleiderer,
\newblock \emph{Phys. Rev. B} \textbf{2010}, \emph{81} 041203(R).

\bibitem{Bau16}
A.~Bauer, M.~Garst, C.~Pfleiderer,
\newblock \emph{Phys. Rev. B} \textbf{2016}, \emph{93} 235144.

\bibitem{Kag17}
F.~Kagawa, H.~Oike, W.~Koshibae, A.~Kikkawa, Y.~Okamura, Y.~Taguchi,
  N.~Nagaosa, Y.~Tokura,
\newblock \emph{Nat. Commun.} \textbf{2017}, \emph{8} 1332.

\bibitem{Bau18}
A.~Bauer, A.~Chacon, M.~Halder, C.~Pfleiderer,
\newblock \emph{Skyrmion Lattices Far from Equilibrium}, 151--176,
\newblock Springer International Publishing, Cham,
\newblock ISBN 978-3-319-97334-0, \textbf{2018}.

\bibitem{Lit24}
M.~T. Littlehales, S.~H. Moody, L.~A. Turnbull, B.~M. Huddart, B.~A. Brereton,
  G.~Balakrishnan, R.~Fan, P.~Steadman, P.~D. Hatton, M.~N. Wilson,
\newblock \emph{Nano Lett.} \textbf{2024}, \emph{24} 6813.

\bibitem{Kal24}
J.~Kalin, S.~Sievers, H.~Schumacher, R.~Abram, H.~F\"user, M.~Bieler, D.~Kalin,
  A.~Bauer, C.~Pfleiderer,
\newblock \emph{Phys. Rev. Appl.} \textbf{2024}, \emph{21} 034065.

\bibitem{Oka16}
Y.~Okamura, F.~Kagawa, S.~Seki, Y.~Tokura,
\newblock \emph{Nat. Commun.} \textbf{2016}, \emph{7} 12669.

\bibitem{Mak17}
K.~Makino, J.~D. Reim, D.~Higashi, D.~Okuyama, T.~J. Sato, Y.~Nambu, E.~P.
  Gilbert, N.~Booth, S.~Seki, Y.~Tokura,
\newblock \emph{Phys. Rev. B} \textbf{2017}, \emph{95} 134412.

\bibitem{Bir21}
M.~T. Birch, D.~Cort\'{e}s-Ortu\~{n}o, N.~D. Khanh, S.~Seki,
  A.~\v{S}tefan\v{c}i\v{c}, G.~Balakrishnan, Y.~Tokura, P.~D. Hatton,
\newblock \emph{Communications Physics} \textbf{2021}, \emph{4} 175.

\bibitem{Kar17}
K.~Karube, J.~S. White, D.~Morikawa, M.~Bartkowiak, A.~Kikkawa, Y.~Tokunaga,
  T.~Arima, H.~M. R\o{}nnow, Y.~Tokura, Y.~Taguchi,
\newblock \emph{Phys. Rev. Materials} \textbf{2017}, \emph{1} 074405.

\bibitem{Muh09}
S.~M\"{u}hlbauer, B.~Binz, F.~Jonietz, C.~Pfleiderer, A.~Rosch, A.~Neubauer,
  R.~Georgii, P.~B\"{o}ni,
\newblock \emph{Science} \textbf{2009}, \emph{323} 915.

\bibitem{Kru18}
A.~J. Kruchkov, J.~S. White, M.~Bartkowiak, I.~\v{Z}ivkovi\'{c}, A.~Magrez,
  H.~M. R{\o}nnow,
\newblock \emph{Sci. Rep.} \textbf{2018}, \emph{8} 10466.

\bibitem{Mil13}
P.~Milde, D.~K\"{o}hler, J.~Seidel, L.~M. Eng, A.~Bauer, A.~Chacon,
  J.~Kindervater, S.~M\"{u}hlbauer, C.~Pfleiderer, S.~Buhrandt, C.~Sch\"{u}tte,
  A.~Rosch,
\newblock \emph{Science} \textbf{2013}, \emph{340} 1076.

\bibitem{Nak19}
T.~Nakajima, K.~Karube, Y.~Ishikawa, M.~Yonemura, N.~Reynolds, J.~S. White,
  H.~M. R\o{}nnow, A.~Kikkawa, Y.~Tokunaga, Y.~Taguchi, Y.~Tokura, T.~Arima,
\newblock \emph{Phys. Rev. B} \textbf{2019}, \emph{100} 064407.

\bibitem{Boc19}
J.~D. Bocarsly, C.~Heikes, C.~M. Brown, S.~D. Wilson, R.~Seshadri,
\newblock \emph{Phys. Rev. Mater.} \textbf{2019}, \emph{3} 014402.

\bibitem{Kar18}
K.~Karube, J.~S. White, D.~Morikawa, C.~D. Dewhurst, R.~Cubitt, A.~Kikkawa,
  X.-Z. Yu, Y.~Tokunaga, T.~Arima, H.~M. R\o~nnow, Y.~Tokura, Y.~Taguchi,
\newblock \emph{Sci. Adv.} \textbf{2018}, \emph{4} aar7043.

\bibitem{Pre21}
M.~Prei\ss{}inger, K.~Karube, D.~Ehlers, B.~Szigeti, H.-A. Krug~von Nidda,
  J.~S. White, V.~Ukleev, H.~M. R\o{}nnow, Y.~Tokunaga, A.~Kikkawa, Y.~Tokura,
  Y.~Taguchi, I.~K\'{e}zsm\'{a}rki,
\newblock \emph{npj Quantum Materials} \textbf{2021}, \emph{6} 65.

\bibitem{Whi22}
J.~S. White, K.~Karube, V.~Ukleev, P.~M. Derlet, R.~Cubitt, C.~D. Dewhurst,
  A.~R. Wildes, X.~Z. Yu, H.~M. R{\o}nnow, Y.~Tokura, Y.~Taguchi,
\newblock \emph{Journal of Applied Crystallography} \textbf{2022}, \emph{55}
  1219.

\bibitem{Kar22}
K.~Karube, Y.~Taguchi,
\newblock \emph{APL Mater.} \textbf{2022}, \emph{10} 080902.

\bibitem{Tok15}
Y.~Tokunaga, X.~Z. Yu, J.~S. White, H.~R\o~nnow, D.~Morikawa, Y.~Taguchi,
  Y.~Tokura,
\newblock \emph{Nat. Commun.} \textbf{2015}, \emph{6} 7638.

\bibitem{Ukl19}
V.~Ukleev, Y.~Yamasaki, D.~Morikawa, K.~Karube, K.~Shibata, Y.~Tokunaga,
  Y.~Okamura, K.~Amemiya, M.~Valvidares, H.~Nakao, Y.~Taguchi, Y.~Tokura,
  T.~Arima,
\newblock \emph{Phys. Rev. B} \textbf{2019}, \emph{99} 144408.

\bibitem{Ukl21}
V.~Ukleev, K.~Karube, P.~M. Derlet, C.~N. Wang, H.~Luetkens, D.~Morikawa,
  A.~Kikkawa, L.~Mangin-Thro, A.~R. Wildes, Y.~Yamasaki, Y.~Yokoyama, L.~Yu,
  C.~Piamonteze, N.~Jaouen, Y.~Tokunaga, H.~M. R{\o}nnow, T.~Arima, Y.~Tokura,
  Y.~Taguchi, J.~S. White,
\newblock \emph{npj Quantum Materials} \textbf{2021}, \emph{6} 40.

\bibitem{Yu18}
X.~Z. Yu, W.~Koshibae, Y.~Tokunaga, K.~Shibata, Y.~Taguchi, N.~Nagaosa,
  Y.~Tokura,
\newblock \emph{Nature} \textbf{2018}, \emph{564} 95.

\bibitem{Yi09}
S.~D. Yi, S.~Onoda, N.~Nagaosa, J.~H. Han,
\newblock \emph{Phys. Rev. B} \textbf{2009}, \emph{80} 054416.

\bibitem{Lin15}
S.-Z. Lin, A.~Saxena, C.~D. Batista,
\newblock \emph{Phys. Rev. B} \textbf{2015}, \emph{91} 224407.

\bibitem{Pup20}
P.~Puphal, V.~Pomjakushin, N.~Kanazawa, V.~Ukleev, D.~J. Gawryluk, J.~Ma,
  M.~Naamneh, N.~C. Plumb, L.~Keller, R.~Cubitt, E.~Pomjakushina, J.~S. White,
\newblock \emph{Phys. Rev. Lett.} \textbf{2020}, \emph{124} 017202.

\bibitem{Hay21}
S.~Hayami, Y.~Motome,
\newblock \emph{J. Phys.: Condens. Matter} \textbf{2021}, \emph{33} 443001.

\bibitem{Leo24}
A.~O. Leonov,
\newblock \emph{Nanomaterials} \textbf{2024}, \emph{14} 1524.

\bibitem{Hen23}
M.~E. Henderson, B.~Heacock, M.~Bleuel, D.~G. Cory, C.~Heikes, M.~G. Huber,
  J.~Krzywon, O.~Nahman-Levesqué, G.~M. Luke, M.~Pula, D.~Sarenac,
  K.~Zhernenkov, D.~A. Pushin,
\newblock \emph{Nat. Phys.} \textbf{2023}, \emph{19} 1617.

\bibitem{Nee24}
J.~Neethirajan, B.~J. Daurer, M.~D.~P. Mart\'{\i}nez, A.~c.~v. Hrabec,
  L.~Turnbull, R.~Yamamoto, M.~R. Ferreira, A.~c.~v. \ifmmode \check{S}\else
  \v{S}\fi{}tefan\ifmmode \check{c}\else \v{c}\fi{}i\ifmmode~\check{c}\else
  \v{c}\fi{}, D.~A. Mayoh, G.~Balakrishnan, Z.~Pei, P.~Xue, L.~Chang, E.~Ringe,
  R.~Harrison, S.~Valencia, M.~Kazemian, B.~Kaulich, C.~Donnelly,
\newblock \emph{Phys. Rev. X} \textbf{2024}, \emph{14} 031028.

\bibitem{Mor17}
D.~Morikawa, X.~Yu, K.~Karube, Y.~Tokunaga, Y.~Taguchi, T.-h. Arima, Y.~Tokura,
\newblock \emph{Nano Letters} \textbf{2017}, \emph{17} 1637.

\bibitem{Nag19}
T.~Nagase, M.~Komatsu, Y.~G. So, T.~Ishida, H.~Yoshida, Y.~Kawaguchi,
  Y.~Tanaka, K.~Saitoh, N.~Ikarashi, M.~Kuwahara, M.~Nagao,
\newblock \emph{Phys. Rev. Lett.} \textbf{2019}, \emph{123} 137203.

\bibitem{Cri24}
J.~C. Criado, P.~D. Hatton, A.~Lanza, S.~Schenk, M.~Spannowsky,
\newblock \emph{Phys. Rev. B} \textbf{2024}, \emph{109} 195114.

\bibitem{Psa21}
C.~Psaroudaki, C.~Panagopoulos,
\newblock \emph{Phys. Rev. Lett.} \textbf{2021}, \emph{127} 067201.

\bibitem{Dew23}
C.~D. Dewhurst,
\newblock \emph{Journal of Applied Crystallography} \textbf{2023}, \emph{56}
  1595.

\bibitem{Van14}
A.~Vansteenkiste, J.~Leliaert, M.~Dvornik, M.~Helsen, F.~Garcia-Sanchez,
  B.~Van~Waeyenberge,
\newblock \emph{AIP advances} \textbf{2014}, \emph{4}, 10.

\bibitem{Hen22}
M.~E. Henderson, M.~Bleuel, J.~Beare, D.~G. Cory, B.~Heacock, M.~G. Huber,
  G.~M. Luke, M.~Pula, D.~Sarenac, S.~Sharma, E.~M. Smith, K.~Zhernenkov, D.~A.
  Pushin,
\newblock \emph{Phys. Rev. B} \textbf{2022}, \emph{106} 094435.

\end{thebibliography}




%
%
%
%


%

\end{document}